\documentclass{aa}  

\usepackage{graphicx}
\usepackage{txfonts}
\usepackage{xcolor}
\AddToHook{begindocument/before}{\usepackage{hyperref}}
\newcommand{\Msun}{\ensuremath{\mathrm{M}_\odot}}
\newcommand{\Te}{\ensuremath{T_\mathrm{e}}}
\newcommand{\Ne}{\ensuremath{N_\mathrm{e}}}
\newcommand{\NH}{\ensuremath{N_\mathrm{H}}}
\newcommand{\Ioni}{\ensuremath{X^{+i}}}
\newcommand{\TeMet}{{\Te-method}}
\newcommand{\Kelv}{\ensuremath{\,\mathrm{K}}}
\newcommand{\Ha}{\ensuremath{\mathrm{H}\alpha}}
\newcommand{\Hb}{\ensuremath{\mathrm{H}\beta}}
\newcommand{\SIII}{\ensuremath{\mathrm{[S\,\textsc{iii}]}}}
\newcommand{\SII}{\ensuremath{\mathrm{[S\,\textsc{ii}]}}}
\newcommand{\OIII}{\ensuremath{\mathrm{[O\,\textsc{iii}]}}}
\newcommand{\OII}{\ensuremath{\mathrm{[O\,\textsc{ii}]}}}
\newcommand{\OI}{\ensuremath{\mathrm{[O\,\textsc{i}]}}}
\newcommand{\NII}{\ensuremath{\mathrm{[N\,\textsc{ii}]}}}
\newcommand{\HII}{\ensuremath{\mathrm{H\,\textsc{ii}}}}
\newcommand{\LF}{\ensuremath{\mathcal{L}}}

\begin{document}

   \title{HOMERUN\thanks{Highly Optimized Multi-cloud Emission-line Ratios Using photo-ionizatioN} a new approach to photoionization modelling}

   \subtitle{I - reproducing observed emission lines with percent accuracy and obtaining accurate physical properties of the ionized gas}

   \author{A. Marconi
          \inst{1,2}\fnmsep\thanks{\email{alessandro.marconi@unifi.it}}
          \and
          A. Amiri\inst{3,1,2}
          \and
          A. Feltre\inst{2}
          \and
          F. Belfiore\inst{2}
          \and
          G. Cresci\inst{2}
           \and
          M. Curti\inst{4}
              \and
          F. Mannucci\inst{2}
          %%%%%
          \and
          E. Bertola\inst{2}
          \and
          M. Brazzini\inst{1,2}
          \and
          S. Carniani\inst{7,2}
          \and
          E. Cataldi\inst{1,2}
          \and
          Q. D'Amato\inst{2}
          \and
          G. de Rosa\inst{8}
          \and
          E. Di Teodoro\inst{1}
          \and
          M. Ginolfi\inst{1,2}
          \and
          N. Kumari\inst{8}
          \and
          C. Marconcini\inst{1,2}
          \and
          R. Maiolino\inst{5,6}
          \and
          L. Magrini\inst{2}
          \and
          A. Marasco\inst{9}
          \and
          M. Mingozzi\inst{8}
          \and
          B. Moreschini\inst{1,2}
          \and
         T. Nagao\inst{10}
         \and
         E. Oliva\inst{2}
          \and
          M. Scialpi\inst{1,2}
          \and
          N. Tomicic\inst{11,1,2}
          \and
          G. Tozzi\inst{1,2}
          \and
          L. Ulivi\inst{1,2,12}
          \and
          G. Venturi\inst{7,2}
          }

   \institute{Dipartimento di Fisica e Astronomia, Università di Firenze, Via G. Sansone 1, I-50019, Sesto F.no (Firenze), Italy
         \and
             INAF-Osservatorio Astrofisico di Arcetri, Largo E. Fermi 5, I-50125, Firenze, Italy
         \and
         Department of Physics, University of Arkansas, 226 Physics Building, 825 West Dickson Street, Fayetteville, AR 72701, USA
         \and
         European Southern Observatory, Karl-Schwarzschild-Strasse 2, 85748 Garching, Germany
         \and
            Kavli Institute for Cosmology, University of Cambridge, Madingley Road, Cambridge, CB3 OHA, UK
          \and
             Cavendish Laboratory - Astrophysics Group, University of Cambridge, 19 JJ Thomson Avenue, Cambridge, CB3 OHE, UK
            \and
            Scuola Normale Superiore, Piazza dei Cavalieri 7, I-56126 Pisa, Italy
            \and
            Space Telescope Science Institute, 3700 San Martin Drive. Baltimore, MD, 21210
            \and
            INAF-Padova Astronomical Observatory, Vicolo Osservatorio 5, 35122, Padova, Italy
	        \and Research Center for Space and Cosmic Evolution, Ehime University, Matsuyama, Ehime 790-8577, Japan
            \and Department of Physics, Faculty of Science, University of Zagreb, Bijeni\v{c}ka 32, 10 000 Zagreb, Croatia
            \and Department of Physics, University of Trento, Via Sommarive 14, I-38123 Povo (Trento), Italy
             }

   \date{Received XXXX; accepted XXXXX}

\abstract{We present HOMERUN (Highly Optimized Multi-cloud Emission-line Ratios Using photo-ionizatioN), a new approach to modelling emission lines from photoionized gas that can simultaneously reproduce all observed line intensities from a wide range of ionization levels and with high accuracy. 
Our approach is based on the weighted combination of multiple single-cloud photoionization models and, contrary to previous works, the novelty of our approach consists in using the weights as free parameters of the fit and constraining them with the observed data. 
One of the main applications of HOMERUN is the accurate determination of gas-phase metallicities and 
we show that a critical point is to allow for a variation of the N/O and S/O abundance ratios which can significantly improve the quality of the fit and the accuracy of the results. Moreover, our approach provides a major improvement compared to the single-cloud, constant-pressure models commonly used in the literature.
By using high-quality literature spectra of \HII\ regions where 10 to 20 emission lines (including several auroral lines) are detected with 
high signal-to-noise ratio, we show that all lines are reproduced by the model with an accuracy better than 10\%. In particular, the model is able to simultaneously reproduce \OI${\lambda\lambda 6300, 6363}$, \OII${\lambda\lambda 3726,3729}$, \OIII${\lambda \lambda 4959, 5007}$, \SII${\lambda\lambda 6717,6731}$, and \SIII${\lambda\lambda 9069, 9532}$ emission lines  which, to our knowledge, is an unprecedented result. 
Finally, we show that the gas metallicities estimated with our models for HII regions in the Milky Way are in agreement with the stellar metallicities than the estimates based on the \Te-method. 
Overall, our method provides a new accurate tool to estimate the metallicity and the physical conditions of the ionized gas. It can be applied to many different science cases from HII regions to AGN and wherever there are emission lines from photoionized gas. }

   \maketitle

\section{Introduction\label{sec:introduction}}

Metallicity is a key observable of galaxies as element abundances and their evolution across cosmic epochs provide unique information on the physical processes driving the evolution of galaxies (see, e.g., \citealt{maiolino:2019} for a review). In particular, since all elements heavier than Lithium and up to Iron and Nickel are created in stellar cores, the amount of metals provides a measure of the integrated star formation history of galaxies. Furthermore, the existence of scaling relations between metallicity and galaxy properties such as the mass-metallicity  \citep[e.g.][]{tremonti:2004} and fundamental metallicity relations \citep[e.g.][]{mannucci:2010} also indicates that metallicity traces the infall and outflow of gas to and from galaxies. Metallicity can therefore be used to constrain feedback/fueling processes and growth timescales in galaxies (e.g., \citealt{mannucci:2009,cresci:2010,kewley_understanding_2019-1, kumari_extension_2021, curti_chemical_2023} and references therein).

The metallicity of a galaxy and of the surrounding medium can be measured from a range of observables: from the integrated emission of stars, from the warm ($\Te\sim 10^4\Kelv$) ionized interstellar medium (ISM), from the hot  ($\Te\sim 10^6\Kelv$) ionized intracluster (ICM) and circumgalactic medium (CGM), and from the absorption features caused by cold/warm ICM/CGM on the spectra of distant quasars and galaxies. 
In this paper, we focus on the gas-phase metallicity of the ISM warm ionized gas ($\Te\sim 10^4\Kelv$) mainly located within the HII regions surrounding massive O and B stars but also in the form of diffused ionized gas (DIG). 

The emission line spectrum of photoionized gas can be used to constrain the properties of the ionizing source(s) and provides information on gas physical properties like density, temperature, and metallicity. In particular, the measurement of metallicity depends on the density, temperature, and ionization status of the ionized gas 
(e.g., \citealt{osterbrock_astrophysics_2006, netzer_ionized_2008, stasinska_what_2009, kewley_understanding_2019-1}).
Three methods are usually adopted: two observational ones, based on direct electron temperature determinations and on metal recombination lines, respectively; and a theoretical one based on photoionization models (see, section 3 of \cite{maiolino:2019} and references therein for a detailed description and discussion on these methods).

The first observational method, the so-called \textit{\TeMet}, is based on density and temperature estimates from the ratios of collisionally excited forbidden emission lines. It assumes that the complex stratification of the physical properties in the ionized gas volume can be modeled with a few (2-4) homogeneous regions 
characterized by constant values of electron density (\Ne), temperature (\Te), and ionization fractions of any ion \Ioni; $N_e$ and $T_e$ of each region can then be determined from one or more line ratios sensitive to density, like \SII 6730/6716, \OII 3729/3728, and temperature like \OIII 4363/5007, \OII 7330/3728, \NII 5755/6548, and \SIII 6312/9532 (e.g., \citealt{osterbrock_astrophysics_2006, chaosIV, mingozzi_classy_2022} and references therein).  
The emissivity $J_{k,l}$ of a specific emission line originating from the $k,l$ transition, whether due to collisional excitation or recombination, can be determined using these \Ne\ and \Te values. Subsequently, the relative abundance of the emitting ion can be obtained from the reddening corrected observed ratio between the emission line of \Ioni and a hydrogen line, typically \Hb. Both lines are assumed to be emitted from the same volume $\Delta V$.
In formulae,
\begin{eqnarray}
L_{k,l}(\Ioni) & \simeq & N(\Ioni)\, \Ne\, J_{k,l}(\Ioni; \Ne, \Te)\,\Delta V(\Ioni),\\
L(\Hb)          & \simeq & N(\mathrm{H}^{+})\, \Ne\, J_{\Hb}(\Ne, \Te)\,\Delta V(H\beta),\\
 \frac{N(\Ioni)}{N(\mathrm{H})} & \simeq &\frac{L_{k,l}(\Ioni)}{L(\Hb)} \frac{J_{\Hb}(\Ne, \Te)}{J_{k,l}(\Ioni; \Ne, \Te)},
\end{eqnarray}
where $L$ is a line luminosity and it is assumed that $N(\mathrm{H}^+) \simeq N(\mathrm{H})$ and $\Delta V(\Ioni) \simeq \Delta V(\Hb)$. 
The abundance of an element $X$ can then be determined by adding up the abundances of its ions:
\begin{equation} 
\frac{N(X)}{N(\mathrm{H})} = \sum_i \frac{N(\Ioni)}{N(\mathrm{H})}.
\end {equation}

Reliable values of $T_e$ are needed because of the strong, exponential dependence of $J_{k,l}$ on temperature. Hereafter, we refer to these abundances as \Te-abundances.

The second observational method is based on metal recombination lines and is similar to the aforementioned approach, with the only fundamental difference that the line emissivity $J_{k,l}(X^{+i}; \Ne, \Te)$ has only a weak, roughly linear dependence on temperature whose exact value is therefore much less important than for the \Te-method (e.g. \citealt{peimbert_densities_2013, esteban_carbon_2014, toribio_san_cipriano_carbon_2017}).

Although commonly considered observational methods, both approaches necessarily rely on photoionization calculations to estimate ionic temperatures and abundances not directly accessible from the observations (see, e.g., \citealt{gutkin:2016, amayo_ionization_2021} and references therein). Moreover, abundances derived from the \Te-method are known to disagree with those derived from metal recombination lines. A possible explanation is the existence of temperature fluctuations which mostly affect the determinations based on collisionally excited lines (e.g., \citealt{peimbert:1967, peimbert_densities_2013, toribio_san_cipriano_carbon_2017, nicholls_estimating_2020, mendez-delgado_temperature_2023}). 
Also, especially for distant galaxies, the temperature is usually derived for only one of the zones  of the HII regions and locally calibrated relations are used to extrapolate the temperatures of the other zone(s) \citep[e.g.][]{curti_new_2017}, potentially introducing large uncertainties.

The theoretical method is instead based on a comparison between the observed line ratios and those predicted from photoionization models. In practice, given a set of observed emission lines, a suite of photoionization models (\textit{single-cloud} models) for different ionizing continua and physical properties of the gas (including metallicity) is created to best match the observed emission lines thus determining the gas metallicity among the other physical properties (e.g., \citealt{lopez-sanchez_eliminating_2012, nicholls_resolving_2012, binette_discrepancies_2012, pilyugin_counterpart_2012, dopita_new_2013, blanc:2015, pilyugin_new_2016, vale_asari_bond_2016, perez-montero_using_2017, strom_measuring_2018, mignoli:2019, papovich_clear_2022}). 
Hereafter, we refer to this method as the \emph{Theoretical} method and to the metal abundances derived with it as the \emph{Model} abundances.

Overall, all these methods require the measurement of several emission lines to constrain metallicities. The \Te-method, in particular, requires the measurement of the so-called \emph{auroral} lines, notable examples of which are \SII 4069, \OIII 4363, \OII 7330, \NII 5755, and \SIII 6312, which are needed for temperature determination but are usually quite faint (less than a few percent of the brightest lines). 
Instead, most photoionization models, at least in the extragalactic field, use exclusively the brightest emission lines which are the only ones usually detected: these are lines like \Hb, \Ha, \OIII 5007, \OII 3727, 3729, \NII 6584, \SII 6716, 6730, \SIII 9532\ and are usually referred to as \textit{strong lines}. 

For the sources where it is possible to measure only these strong emission lines, it is  customary to use the so-called \emph{Strong-Line} method, where metallicity is estimated from one or more strong line ratios, whose dependence on metallicity has been calibrated using $T_e$-based and/or $Model$-based metallicities from HII-regions and galaxies with high S/N spectra (e.g., \citealt{kewley_using_2002, pettini_oiiinii_2004, tremonti:2004, kewley:2008, lopez-sanchez_eliminating_2012, dopita_new_2013, curti_new_2017, kewley_understanding_2019-1, sanders_direct_2023}). These line ratios have clearly defined metallicity dependencies because of existing correlations between the physical parameters of gas clouds, like the anti-correlation between ionization parameter and metallicity  (see, e.g., \citealt{carton:2017, Ji:2022}) or star formation rate (e.g., \citealt{papovich_clear_2022}). 
These methods, and especially the \emph{Strong-Line} one, have extensively been used in studies of galaxy evolution, 
providing evidence for the well-known metallicity scaling relations mentioned above.

However, there are still important open issues in metallicity estimates: contradicting calibrations of the \emph{Strong-Line} method exist in the literature. These disagreements are due to the different reference HII or galaxy samples used by the different authors, the differences in the application of the $T_e$ method (i.e. corrections for temperature fluctuations, ionization corrections), and the different assumptions used for the photoionization models (abundance ratios, ionizing continua, dust properties etc.). Moreover, the $T_e$ method and \emph{Theoretical} methods based on photoionization models usually provide discrepant results, with the latter suggesting metallicities which can be larger by 0.3 dex or more, leading to significant uncertainties on all scaling relations involving metallicity (e.g., \citealt{kewley:2008, stasinska_nebular_2010, lopez-sanchez_eliminating_2012, curti_mass-metallicity_2020}).  

In general, it is clear that both approaches aim to reproduce the complexity of ionized clouds under oversimplified assumptions: a single HII region is a complex structure made of condensations and filaments of gas with non-trivial gradients of physical properties, possibly exposed to the ionizing photons emitted by different sources, and the emission line spectrum of a star-forming galaxy is the superposition of several hundreds of HII regions, each with different physical conditions (e.g. \citealt{charlot:2000}).  

Overall, the recent important advances in spectroscopic instrumentation, in terms of both quality and quantity of the data brought forward by integral field spectrographs, have not been matched by similar improvements in the methods used for metallicity measurements. On one hand, the $T_e$-method has not changed in decades, apart from (semi)empirical corrections or improved theoretical ionization corrections (e.g., \citealt{mendez-delgado_temperature_2023}. On the other hand, the physics within photoionization models has been greatly improved, but the modeling of the observed data is still mostly based on single-cloud models or on simple combinations of multiple clouds: there have been a few attempts to model the complexity of real gaseous nebulae using, e.g., Montecarlo techniques (\citealt{ercolano_mocassin_2003, jin_messenger_2022}) but at the expense of the accuracy of the physical processes included in codes.
Still, the successes and improvements of photoionization models over the last years indicate that they could represent the way forward to improve metallicity estimates.

Here we present HOMERUN (Highly Optimized Multi-cloud Emission-line Ratios Using photo-ionizatioN) a new approach to photoionization modelling which builds upon a grid of \textit{single-cloud} photoionization models but, in contrast to other models presented in the literature, combines them with a weighting function which is determined from the observations, not assumed \textit{a-priori}. 
In section \ref{sec:photomodels} we describe the current approaches to photoionization modeling while in \ref{sec:model} we present our new HOMERUN model. In sections \ref{sec:example} through \ref{sec:berg}, we describe and discuss its application to the spectra of samples of HII regions. Finally, in section \ref{sec:milyway} we show how our model resolves the apparent discrepancy between the observed metallicities of HII regions and stars in the Milky Way.
We summarise our findings and draw our conclusions in section \ref{sec:summary}.

\section{Photoionization Models\label{sec:photomodels}}

There are broadly two kinds of approaches in photoionization modelling of HII regions, and in both cases the publicly available CLOUDY\footnote{\href{https://nublado.org}{https://nublado.org}} \citep{ferland:1993, ferland:2017} and MAPPINGS\footnote{\href{https://mappings.readthedocs.io/en/latest/index.html}{https://mappings.readthedocs.io/en/latest/index.html}} \citep{sutherland:1993,dopita:1996} codes are widely used.
One approach is based on comparing single-cloud constant-pressure models with observations (e.g., \citealt{dopita:2016}), also taking into account the star-formation history (e.g., \citealp{charlot:2001, chevallard:2018}).
The other is based on the combination of different single-cloud constant-density models through an assumed weighting function: this is for instance the case of the so-called Locally Optimally-emitting Clouds (LOC; \citealt{baldwin:1995, korista:1998, nagao_evolution_2006}).
The former approach is extensively used in the analysis of galaxy spectra and in the determination of metal abundances in their ISM, while the latter approaches are mostly used in the analysis of the emission lines from the Broad Line Regions of Active Galactic Nuclei (AGN) although it has also been applied to the spectra of star-forming galaxies \citep{richardson_interpreting_2016-1}.
These models are successful in reproducing the general emission-line properties of populations of galaxies and AGN but they struggle in reproducing the emission-line spectra of single objects: different line ratios in the same galaxy are often reproduced by models with different input parameters because the models fail to reproduce the observed wide range of excitation levels which is manifested by the existence of strong emission lines from different ionization states, like \OIII$\lambda$5007, \OII$\lambda$3727 and \OI$\lambda$6300 (e.g., \citealt{Martin:1997, Iglesias-Paramo:2002, Stasinska:2015, Ramambason:2020, Lebouteiller:2022}). 
More in detail, a single-cloud model is described by a slab of gas in plane-parallel or spherical geometry, ionized by a central source. A single-cloud model therefore is defined by: 
\begin{itemize}
\item the spectrum of the ionizing continuum, that can vary widely depending on the model used for the ionizing source (stellar age, stellar and gas-phase metallicity, presence of binary stars, etc.);
\item the hydrogen density \NH\ at the illuminated face of the cloud; 
\item  the radial distance $r_{in}$ of the illuminated face of the cloud from the ionizing source;
\item the ionization parameter $U = Q(\mathrm{H})/(4\pi r_{in}^2\, c\, \NH )$, where $Q(\mathrm{H})$ is the rate of ionizing photons and $c$ the speed of light; sometimes different authors use a different definition of $U$ which is computed, e.g., at a different radius or as a volume average (see for example \citealt{charlot:2001, gutkin:2016, plat:2019}).
\item the abundances of all the elements relative to hydrogen;
\item whether the cloud is at constant pressure or constant density;
\item whether the cloud contains dust, and what are its properties and the fraction of chemical elements depleted into dust;
\item the assumption of stationary equilibrium, i.e. no time evolution;
\item a criterion to define the outer edge of the cloud, which could be a limit on any parameter among hydrogen column density, ionization fraction, and electron temperature. 
\end{itemize}
The model is  integrated starting from the illuminated face of the cloud, numerically solving the equations of ionization equilibrium for all ions, the energy balance equation, and the radiative transfer equation; all other relevant physical processes are also considered (e.g, \citealt{osterbrock_astrophysics_2006, stasinska_what_2009, netzer_ionized_2008}; see volume 3 of Cloudy manual "Hazy" for an exhausting description\footnote{Available with the CLOUDY software at \href{https://nublado.org}{nublado.org}}).
The model predicts, among other things, the fluxes or luminosities of all the considered transitions which can then be compared with observations. The metallicity of the model which provides the best match with the observations is then the estimated gas metallicity. 

The ionization level of the gas at a given point in the cloud is set, among other things, by the value of the ionization parameter at that point, i.e. computed from $Q(\mathrm{H})$ and \NH\ at $r$, where $r$ is the distance from the ionizing source. 
In constant-pressure cloud models, as $r$ increases, the temperature decreases and the density grows, leading to a radial decrease in $U$ which is faster than that predicted by constant-density models. This wider $U$ range allows for a better match with the observed emission lines from different ionization stages, and is the reason why constant-pressure, single cloud models are typically more successful than constant-density ones (e.g., \citealt{dopita:2002, stern_radiation_2014, kewley_theoretical_2019}).

In order to provide a better match to the observed emission lines, combinations of single-cloud models have been considered. One notable example is that of the LOC models mentioned above. Here, a grid of single clouds is computed, each characterized by a single value of  $N_H$ and  $\Phi(\mathrm{H}) = Q(\mathrm{H})/(4\pi r_{in}^2)$ (the ionizing photon flux),  and the emission lines predicted by the single-cloud models are combined with a weighting function which is usually proportional to $\Phi(\mathrm{H})^{-\alpha} \, \NH^{-\beta}$, with $\alpha$ and $\beta$ two slopes of order of unity which can be free parameters of the model \citep{nagao_evolution_2006}. As in the case of single-cloud models, the metallicity of the model which provides the best match is then the estimated gas metallicity. 
A different approach, mostly applied to AGN, assumes that the clouds are confined by radiation or magnetic pressure  and that their covering factor varies as a power law of distance $r$ from the ionizing source, resulting in \NH\ and $U$ which vary similarly as a power law of $r$ (e.g., \citealt{netzer:2020}). LOC models are better suited to reproduce the complexity of a real HII region and its irregular geometrical and ionization structure, but the use of a fixed weighting scheme is still a severe limitation.

\section{The HOMERUN model\label{sec:model}}

\subsection{The basic assumption: a combination of multiple single-cloud models}

The luminosity of an emission line $l$ of ion \Ioni\ from either recombination or collisional excitation can be written as
\begin{equation}\label{eq:linelum}
L_l(\Ioni)_{obs} = \int_V \Ne\, N(\Ioni)\, J_l(\Ioni, \Te)\, \mathrm{d}V,
\end{equation}
where $V$ is the volume of the ionized gas, \Ne\ and $N(\Ioni)$ are the electron and ion densities, respectively, \Te\ is the electron temperature and $J_l(\Ioni, \Te)$ is the emissivity of line $l$. The line emissivity $J_l$ is the same as in Eq. 1 but here we simply indicate the transition as $l$ and we do not explicitly mention the usually weak dependence on the electron density.
$\Ne$, $N(\Ioni)$ and $\Te$ are physical quantities which depend on the position within the volume, yet the luminosity, which is an integrated quantity, does not depend on the detailed density and temperature distribution of the gas but on their weighted average values over the entire gas volume. 
Therefore, any model that  has the correct volume, average density and temperature of the \Ioni\ region, will reproduce $L_l(\Ioni)_{obs}$, even if the spatial distribution of the gas physical properties does not match the real one.  

In this study we assume that the ion emission line luminosity defined in Eq. \ref{eq:linelum} can be sufficiently well approximated by a linear combination of  $m$ single cloud models, each characterized by its own ionisation parameter and density
\begin{eqnarray}
L_l(\Ioni)_{mod} & = & \sum_{j=1}^m\, w_j\, L_l(\Ioni)_{mod,j} \\
                          & = & \sum_{j=1}^m\, w_j \int_{V_j} \Ne\, N(\Ioni) J_l(\Ioni, \Te)\, dV
\end{eqnarray}
where $L_l(\Ioni)_{mod,j}$ is a set of \emph{simple}, single cloud photoionization models which can be considered as the basic building blocks of the model.
The non-negative weights $w_j$ can either be assumed (like in the LOC models) or can be left as free parameters and determined from fitting the observed lines.
The main underlying assumption  is that if a model is able to accurately reproduce the luminosity of many different emission lines from a few different ions, then it has  the correct combinations of volume, average density and temperature for each ion, regardless on their real spatial distribution.
With respect to single cloud models, we expect this approach to provide a more realistic description of the complexity of the warm ionized ISM.

\subsection{The single-cloud models}
The single-cloud models used in the following were computed using CLOUDY v22.01 \citep{ferland:2017} with the following input parameters:
\begin{itemize}
\item BPASS simple stellar population models (v2.3, \citealt{bpass:2022, stanway:2018}) including binary stellar evolution, characterized by a \cite{kroupa:2001} initial mass function from 0.1 to 300 \Msun, [$\alpha$/Fe] = +0.0, metallicities Solar and 0.1 Solar, a single burst of star formation and ages (log yrs) 6, 6.3, 6.7 7.1 (for Solar met.) and 6, 6.4, 6.7, 7.3 (for 0.1 $Z_\odot$)
\item Abundances $Z$ from -2 to +0.4 in steps of 0.2 dex relative to Solar; the Solar abundance pattern is taken from \cite{gass:2010}.
\item Nitrogen is rescaled as a secondary element with respect to oxygen as $[N/O] = -1.5+1.4*([O/H]-[O/H]_{c})$ for $[O/H]\le [O/H]_{crit}$ with $[O/H]_{c}=8.3$ (see, eg., fig. 49 of \citealt{maiolino:2019}); we also consider cases in which $[N/O]$ varies by $\pm 0.6$ dex with respect to this scaling relation.
\item Sulphur scales linearly with oxygen following the Solar $[S/O]$ abundances ratio, but we also consider the case in which $[S/O]$ varies by $\pm 0.3$ dex.
 %\NOTE{ANTONINO: Thus another assumption of the model is that all "components" considered will have the same metallicity, right? I think this should be explicitly mentioned and properly discussed. Why not using a 3D grid including the metallicity in the procedure? The referee may have these kind of concerns.}
\item Gas with and without dust; dust is included with the \texttt{"grains ism Z log"} CLOUDY command line where \texttt{Z} is the log of the abundance relative to Solar; we then use the command \texttt{"metals deplete"} to take into account the depletion of metals into dust (adopted Depletion factors are the default values of Cloudy - see Table 7.8  and references therein of Hazy1 of Cloudy v22.1).
\item A grid in $U$ and $N_H$ with $\log U$ from -5 to -0.5 in steps of 0.5 dex and  $\log\NH$ from 0 to 7 in steps of 1.0 dex. Here, $U$ represents the ionization parameter at the inner face of the cloud. 
\item All single cloud models have constant density.
\item All single cloud models have a plane parallel, open geometry and are radiation bounded with the following integration stopping criteria: maximum H column density $10^{23}\,\mathrm{cm}^{-2}$, minimum temperature $3000$ K, minimum fraction of electrons 0.05.
\end{itemize}
Overall, these choices of parameters require the computation of $2\times4\times25\times 3\times 3 \times 10 \times 8 \times 2 = 134,784$ CLOUDY models. Then, to increase the sampling, the model line luminosities relative to \Hb\ are log-linearly interpolated across metallicity in steps of 0.02 dex.

The choice of  constant density was made in order to  have the simplest possible physical structure of a single cloud model and it does not affect the final results. Indeed, we have verified that constant pressure models can be reproduced by a combination of a constant density models.

The stellar metallicity, which directly affects the shape of the stellar continuum, has not been tied to the metallicity of the gas to allow more freedom to the modeling. We have additionally verified that the fit results are not much sensitive to the stellar continuum shape which cannot then be accurately constrained; this issue is beyond the scope of this paper and will be further investigated elsewhere.

A single-cloud CLOUDY model is characterized by a total \Hb\ surface brightness. Considering the radial symmetry in a spherical model, it is possible to obtain the volume that the cloud must occupy to reproduce the observed luminosity.
Models with low density and low ionization parameters may therefore become too large to be physical or simply to be included in the apertures from which the spectra are obtained. 
It is easy to show that this implies that these low-density, low-ionization-parameter models have an \Hb\ surface brightness which is too low compared to the observed one. The CHAOS HII regions spectra described below (Sect. \ref{sec:berg}) are obtained through apertures (e.g., slits) whose linear size is $\sim 10\arcsec$ and the  observed \Hb\ surface brightness is, on average, of the order of $\sim 10^{-4}\, \mathrm{erg}\,\mathrm{cm}^{-2}\,\mathrm{s}^{-1}\,\mathrm{arcsec}^{-2}$. Therefore, in this paper, we have  considered $\sim 10^{-5} \mathrm{erg}\,\mathrm{cm}^{-2}\,\mathrm{s}^{-1}\,\mathrm{arcsec}^{-2}$ as the minimum surface brightness that a single-cloud model must have to be retained in the computations. 
This ensures that single-cloud models with weights larger than 10\% of the total account for at least 10\% of the observed surface brightness. The effect of this choice is to discard {between 15 and 20\% of the single-cloud} models in the low density, low ionization parameter corners of the $U,\NH$ grids computed for a given abundance $Z$ and ionizing continuum, {this fraction increasing with increasing metallicity}.
In principle, one should adopt an iterative procedure:  fit the data  - as described below - with all single-cloud models in a grid and find the model weights, compute the surface brightness of the single cloud models given their weights and discard the models with surface brightness lower than the observed surface brightness multiplied by the model weight, redo the fit and iterate until no single cloud model is discarded.

\subsection{The loss function \LF}

The agreement of a multi-cloud model with the observations is quantified with the loss function \LF\  defined below. We consider a grid of single-cloud constant-density, radiation-bound models in ionization parameter $U$ and H-density  \NH\ characterized by
\begin{itemize}
\item the spectrum of the ionizing source: $\mathcal{S}_\nu$
\item the gas metallicity, parameterised with the Oxygen abundance: $\mathcal{A}_O$
\item the relative abundances of metal elements, $\mathcal{Z}$;
\end{itemize}
We identify this grid of models with $\mathcal{G} = \mathcal{G}(\mathcal{S}_\nu, \mathcal{A}_O, \mathcal{Z})$.
Each single-cloud model of the grid $\mathcal{G}$ predicts {the observed} emission line luminosities which we normalise to an arbitrary dimensionless H$\beta$ luminosity of, e.g., 100. {There are no limitations to the emission lines that can be considered in the analysis provided that they are computed by CLOUDY.}
All emission lines are then combined with weights which are found by minimising the loss function (\LF)
\begin{equation}
\LF  = \frac{1}{n} \sum_{i=1}^{n} \left(\frac{L_l(\Ioni)_{obs} -L_l(\Ioni; w_1, w_2, \dots, w_m)_{mod}}{\Delta L_l(\Ioni)_{obs} }\right)^2
\end{equation}
where $n$ is the number of observed emission lines and $m$ is the numbers of single cloud models in the grid. $\Delta L_l(\Ioni)_{obs}$ can be either the  error on the observed emission line luminosity or what we arbitrarily consider an \emph{acceptable} discrepancy between the observed and model luminosities: in the following, we will consider $\Delta L_l(\Ioni)_{obs}$ as the maximum of the two. 

The values of the loss function can be approximately interpreted by assuming that the average relative discrepancy between models and observations is $\varepsilon$ and that for all emission lines and all ions
\begin{eqnarray}
L_{l,mod} & = & L_{l,obs} (1+\varepsilon_r\,\delta_l),\\
\Delta L_{l,obs} & = & \varepsilon_a\, L_{l,obs},
\end{eqnarray}
where $\delta_l$ are random numbers extracted from a normal distribution with zero average, and unit standard deviation. We have distinguished between ``real'' relative discrepancy $\varepsilon_r$ between the models and the observations, and the acceptable discrepancy $\varepsilon_a$, i.e. the one we adopt when computing the loss function. Therefore the loss function is given by
\begin{equation}
\label{eq:discrepancy}
\LF = \frac{1}{n}\sum_l \left(\frac{\varepsilon_r \,\delta_l}{\varepsilon_a }   \right)^2.\end{equation}
For example, if the acceptable discrepancy is equal to the real average discrepancy, the loss function is $\LF \simeq 1.0$. Conversely, assuming an acceptable discrepancy $\varepsilon_a = 0.1$, real average discrepancies of $\varepsilon_r = 0.01, 0.05, 0.1, 0.11, 0.15, 0.2$ will correspond to loss function values of $\LF \simeq 0.01, 0.25, 1.0, 1.2, 2.3, 4.0$, respectively. This numbers should serve as a guideline to estimate the real average discrepancy for any given value of the loss function.

\subsection{Varying the N/O and S/O abundance ratios
}

Usually, the relative abundances of all elements relative to Oxygen are constant and fixed to some predetermined value, like the Solar one. However, in our model fitting we will allow for a scaling of the relative abundances of the elements: 
valid especially for N and S, which have the strongest observed lines in the optical, this choice stems from the evidence presented in the literature, 
and is physically expected given the wide range of conditions in which stars and galaxies form and evolve 
\citep{steidel:2016,morisset_photoionization_2016-2,strom_measuring_2018,maiolino:2019,berg:2020}.
  \begin{figure*}
   \centering
   \includegraphics[width=1\linewidth]{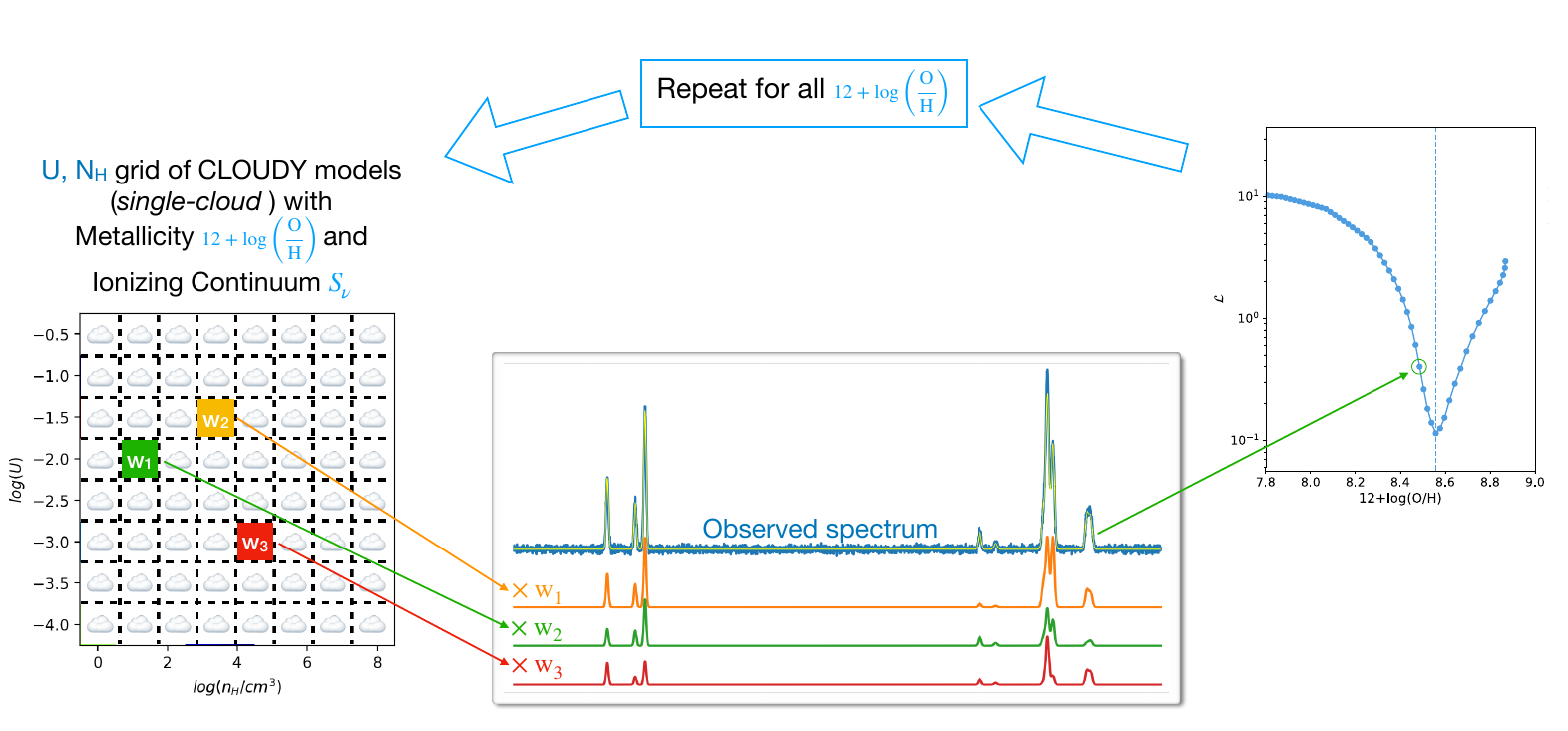}
      \caption{A schematic representation of the fitting procedure: for any $U,\NH$ grid of single-cloud models, we compute the weighted combination of clouds in the grid that gives the best match with the observed lines. Usually only a small fraction of the clouds in the grid have non-zero weights. The goodness of the fit is then quantified with a value of the loss function \LF. The procedure is repeated for all $U,\NH$ grids computed with different gas metallicities and the behaviour of \LF\ as a function of metallicity is obtained. The metallicity of the observed spectrum is identified by the minimum value of \LF. This procedure can be repeated with different ionizing continua thus estimating the effect of varying the ionizing continuum on the metallicity estimate. 
      } 
              
         \label{fig:diagram}
   \end{figure*}

In order to allow for different relative abundances other than the Solar ones used in the computation of the single cloud models, we rescale each set of model emission lines of a given element by a scaling factor which is a free parameter of the fit. 
The only element for which we do not apply such rescaling is oxygen since some of its forbidden emission lines provide the main cooling mechanism of the ionized zone and therefore the line luminosities are not trivially proportional to abundance. Considering, for instance, observed lines from oxygen, N, and S the loss function becomes
\begin{eqnarray}\label{eq:NS}
\LF & = & \sum_{i=1}^{n_{O,H}} \left(\frac{L_l(\Ioni)_{obs} -L_l(\Ioni; \mathbf{w})_{mod}}{\Delta L_l(\Ioni)_{obs} }\right)^2\nonumber\\
& + & \sum_{i=1}^{n_{N}} \left(\frac{L_l(\Ioni)_{obs} -\mathcal{N}*L_l(\Ioni; \mathbf{w})_{mod}}{\Delta L_l(\Ioni)_{obs} }\right)^2\nonumber\\
& + &  \sum_{i=1}^{n_{S}} \left(\frac{L_l(\Ioni)_{obs} -\mathcal{S}*L_l(\Ioni; \mathbf{w})_{mod}}{\Delta L_l(\Ioni)_{obs} }\right)^2
\end{eqnarray}
where $\textbf{w}= (w_1, w_2, \dots, w_m)$ are the weights of the $m$ single-cloud models, $n_{O,H}$ indicates that the sum is made on all $H$ and O lines while $n_{N}$ and $n_{S}$ indicate summation on N and S lines respectively, and $\mathcal{N}$, $\mathcal{S}$ are the scaling factors which are free parameters of the fit.

This procedure assumes that the N and S emission lines scale linearly for small variations of the abundances of these elements. This is not always true, but we have verified that if the abundance of N is varied by less than $\pm 0.3$ dex with respect to the solar value, and S by less than $\pm 0.15$ dex, the line luminosities scale linearly to better than $\sim 10\%$.

Therefore, to correct for the non-linearity of the N and S line intensities with the elements' abundances,  we consider only the models where the scaling factors $\mathcal{N}$ and $\mathcal{S}$ are smaller than $\pm 0.3$ and $\pm 0.15$ dex, respectively. With the choices of $[N/O]$ and $[S/O]$ adopted in the computation of the CLOUDY models, this ensures the accuracy of our computations when varying $[N/O]$ and $[S/O]$ within $\pm 0.9$ and $\pm 0.45$ dex of the Solar values, respectively.

\subsection{The fitting procedure}

In the fitting procedure we consider first a $U, \NH$ grid of single-cloud models computed with a given ionizing continuum and set of element abundances, $\mathcal{G}(\mathcal{S}_\nu, \mathcal{A}_O, \mathcal{Z})$, and find the multi-cloud model which provides the minimum value of the loss function  $\LF_{min}(\mathcal{S}_\nu, \mathcal{A}_O, \mathcal{Z})$. For a given set of scaling factors $\mathcal{N,S}$, the best weights $(w_1, w_2, \dots, w_m)$  are found by solving the non-negative least square (NNLS) problem, in a similar way as in the \texttt{pPXF} python procedure by \cite{cappellari:2004} and by \cite{cappellari:2017}.
Indeed, this model fitting is similar in principle to the fitting of the stellar continua of galaxies where many hundreds or thousands of stellar continuum templates are combined with non-negative weights. Here, emission-line \emph{templates} from the single-cloud models are combined with non-negative weights.
Since in the single-cloud models the emission lines are normalized such that $L$(\Hb)=100, the sum of the weights will be equal to the total observed \Hb\ luminosity. Finally, in finding the best weights we do not apply any regularization, a standard mathematical procedure to mitigate the high-frequency variations in the solutions (e.g. \citealt{cappellari:2017}). 

We repeat for all $U, \NH$ grids of single-cloud models available, iterating over abundances and ionizing continua and, finally, we find the minimum of all $\LF_{min}$ values, thus identifying the  combination of $\mathcal{S}_\nu, \mathcal{A}_O, \mathcal{Z}$ which provides the best fit to the observed line ratios. A schematic representation of the fitting procedure is presented in Fig. \ref{fig:diagram}.

  \begin{figure*}
   \centering
   \includegraphics[width=0.99\linewidth]{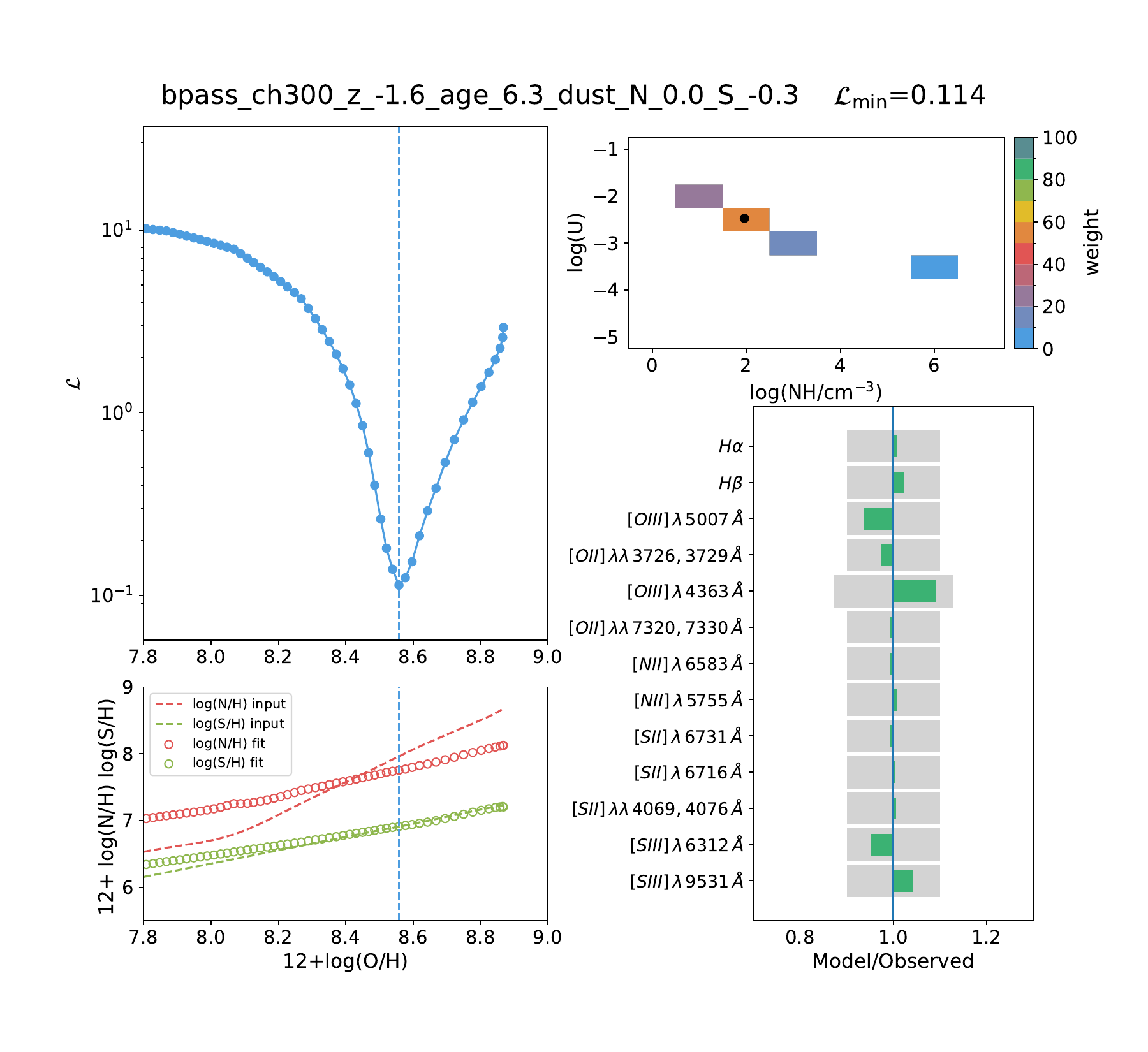}
      \caption{Top left panel: variation of the loss function \LF\ as a function of the Oxygen abundance, $12+\log(O/H)$, for the emission line spectrum of the test HII region. In  this scale the Solar metallicity is 8.69. The blue vertical dashed line represents the metallicity of the model with the minimum value of \LF. 
 Bottom left panel: variation of the N and S abundances  as a function of the Oxygen one. The dashed lines represent the values adopted for the computation of the photoionization models, while the filled circles correspond to the abundances rescaled as described in the text.
 Top right panel: grid of single cloud models in $\log(U)$ and $\log(\NH)$. The colors represent the weights of each single cloud model, as indicated by the colorbar,  when \LF\ reaches the minimum value. In this case only 4 out of 72 models of the grid have non zero weights. The black dot represents the weighted density and ionization parameter of the single cloud models.
Bottom right panel: ratio between  model and observed line luminosities. The gray areas represent the acceptable discrepancy of $\pm 10\%$ or, when larger, the relative error on the observed value.
      } 
              
         \label{fig:model_example}
   \end{figure*}

\subsection{An example of a model fit\label{sec:example}}

We describe the model by presenting an example of its application to the emission line fluxes of an HII region.  The test HII region is randomly extracted from the sample presented by \cite{zurita:2021} and it has all the emission lines listed in the bottom right panel of Figure \ref{fig:model_example} measured with Signal-to-Noise ratio (S/N) larger than 5. 
The results from the application of the model to the entire \cite{zurita:2021} sample of HII regions will be presented in section \ref{sec:zurita}.

We fit the observed lines with the HOMERUN model using as ionizing continuum a BPASS model with age $10^6$ yr and Solar stellar metallicity (Section \ref{sec:model}).
As mentioned in the previous section, the nitrogen lines (\NII$\lambda$5755, \NII$\lambda$6583) and sulphur lines (\SII$\lambda$$\lambda$4069,4076, \SII$\lambda$$\lambda$6716,6731], \SIII$\lambda$6312, \SIII$\lambda$9531) predicted by the model are rescaled by the factors $\mathcal{N}$ and $\mathcal{S}$ which are free parameters of the fit and which take into account that the relative N/O and S/O abundances may be different than those assumed in the photoionization calculations. The same rescaling is not applied to the oxygen lines (\OIII$\lambda$5007, \OIII$\lambda$4363, \OII$\lambda$$\lambda$3726,3729, \OII$\lambda$$\lambda$7320,7330) since they do not depend linearly on abundances. 
For the grids defined by a specific ionizing continuum and gas metallicity value, we find the weights $w_i$ and $\mathcal{N}, \mathcal{S}$ factors which minimise the loss function \LF; finally we select the model with the minimum \LF\ across all metallicities considered. 

The top left panel of Fig. \ref{fig:model_example} shows the variation of the loss function as a function of the gas-phase metallicity, i.e. after dust depletions have been taken into account; the blue vertical dashed line represents the estimated model metallicity, which trivially corresponds to that of the model minimizing \LF. For all lines, we have adopted an \emph{acceptable discrepancy} of 10\% ($\Delta L_l(\Ioni)_{obs}/L_l(\Ioni)_{obs} = 0.1$) when computing \LF.

The bottom left panel of Figure \ref{fig:model_example} represents the nitrogen and sulphur metallicities provided as input in the CLOUDY simulations (dashed coloured lines) and derived from the fit after correcting the input model values with the $\mathcal{N}, \mathcal{S}$  factors (filled circles). The values derived from the fit are close to the relative metal abundances assumed in the Cloudy models.  

The top right panel of Figure \ref{fig:model_example} shows the weights of the single cloud models with different $U$ and \NH\ which were derived for the best fit:  four single clouds out of the 72 in the $U,\NH$ grid have non-zero weights, three of them account for more  than 90\% of the total \Hb\ flux.
Using the weights, it is possible to compute the average density and ionization parameter of  the clouds, which are identified by the black filled circle in top right panel of Figure \ref{fig:model_example}. 

While it is not possible to affirm that the single clouds used in the model have a direct correspondence to the real $N_H,U$ distribution, the average density and ionization parameter derived from the fit are representative of the average density and ionization parameter of the HII region.

In the bottom right panel of Figure \ref{fig:model_example}, we compare the model predictions with the observations: the x-axis represents the ratio between model and  observed line fluxes 
for a certain line, which is indicated on the y-axis. The length of the gray bar represents a $\pm10\%$ discrepancy. In the case of the \OIII 4363 line the gray bar reflects the relative error of the observations which is larger than the adopted 10\% discrepancy.
Finally, the green color indicates a ratio which is within 0.9 and 1.1 (10\% discrepancy max), while orange bars (when present) indicate values outside that range. 

As discussed in section \ref{sec:model}, a loss function value of 0.11, which is the minimum value in the top left panel, can be interpreted as an average discrepancy of $\sim 3.3\%$ between observations and models which is visually confirmed in Figure \ref{fig:model_example}. 

The model is able to reproduce a large number of emission lines from different ionization stages within 5\% discrepancy (with only the \OIII\ 4363 line reproduced with a 10\% discrepancy but still consistent with the observational error): this is a definite improvement with respect to previous models. 
\SIII\ lines are in very good agreement with observations, at variance with previous work in which \SIII\ lines are usually so difficult to reproduce that problems with their atomic parameters have been suggested \citep{mingozzi:2020}. 
The physical parameters of the HII region inferred from the model, such as metallicity {${12+\log(O/H)=8.56}$}, average ionization parameter ${U=-2.47}$ and hydrogen density ${N_H=1.96}$, are also realistic and in agreement with our expectations for HII regions. 
Finally, the plot of the loss function in Figure \ref{fig:model_example} presents a minimum so deep to require a logarithmic scale and indicates that the problem is well defined and not affected by degeneracies, as one might naively expect given the freedom in the choice of the weights. In other words, this model has potentially many parameters (the weights of each single  CLOUDY model) but it is built in  such a way that if the metallicity is not the right one, the lines  can not be well reproduced.

\section{The sample  of HII regions by Zurita et al.\label{sec:zurita}}

\begin{figure*}

  \begin{minipage}[!t]{0.35\textwidth}
    %\centering
    \includegraphics[width=\linewidth]{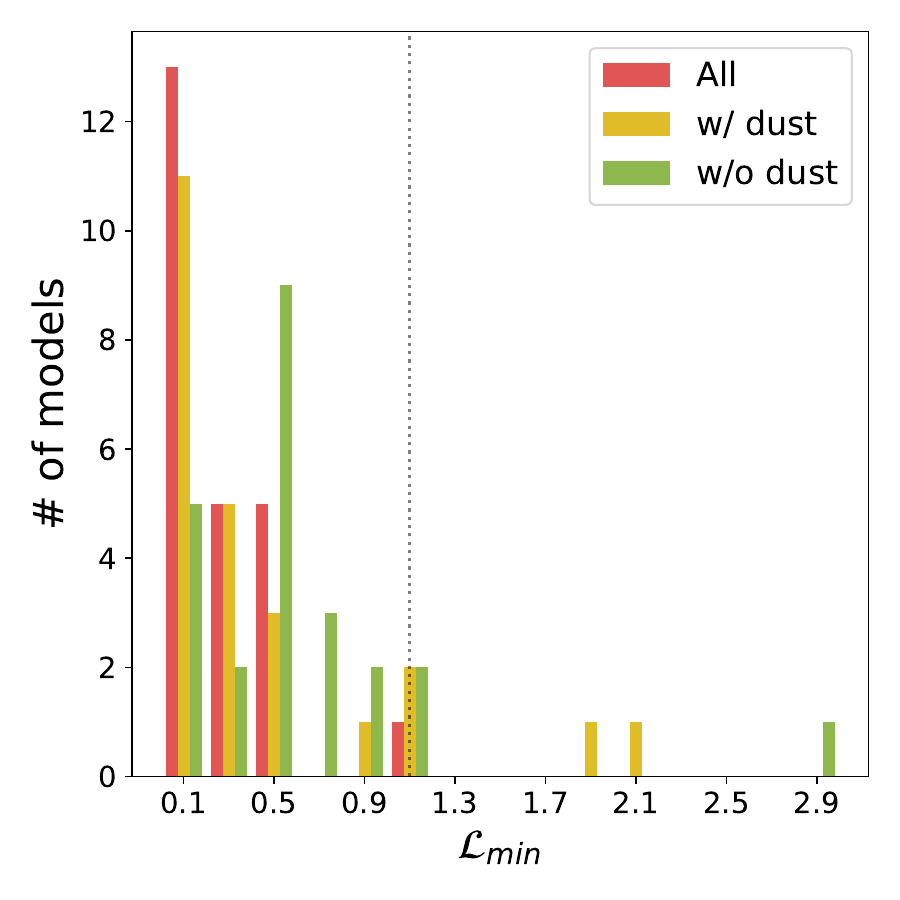}
  \end{minipage}
  \begin{minipage}[!t]{0.64\textwidth}
    %\centering
    \includegraphics[width=\linewidth]{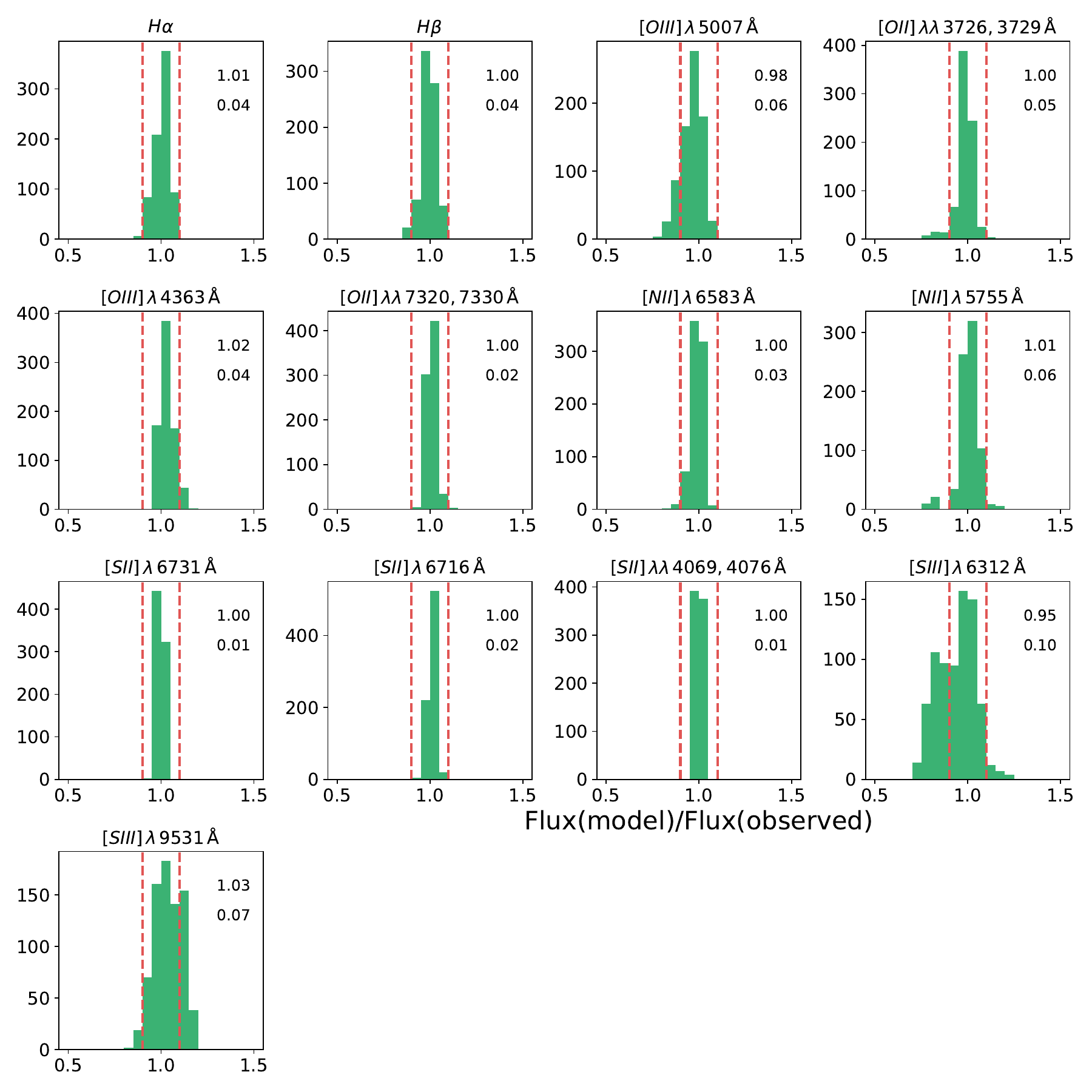}
  \end{minipage}

      \caption{Left: distribution of $\LF_{min}$ values for the subsample of HII regions selected from  \cite{zurita:2021}. Right: distributions of the ratios between model and observed line fluxes for all the selected HII regions and all the models with $\LF\le \LF_{min}+0.25$. The numbers in the top right corner of  the panels represent, from top to bottom, median and standard deviation of distribution of \LF\ values. The vertical dashed lines represent the $\pm 10\%$ discrepancy and show how the large majority of models reproduce all lines within $ \pm 10\%$ of the observed values. }
         \label{fig:zurita_chisq}
\end{figure*}

After presenting the model and its characteristics with a few selected examples, we present its application to larger samples of emission-line sources. In this paper, we focus on HII regions within the Milky Way or local spiral galaxies.

We consider the sample  by \cite{zurita:2021}, who collected the emission line fluxes of 2831 published HII regions from 51 nearby galaxies to study how the presence of bars affects metallicity gradients. The compiled line fluxes, reddening corrected and normalised to \Hb,
include both bright emission lines (\OII$\lambda$$\lambda$3726,3729, \OIII$\lambda$$\lambda$4959,5007, \NII$\lambda$$\lambda$6548,6583, \Ha, \SII$\lambda$$\lambda$6717,6731, \SIII$\lambda$$\lambda$9069,9532) and auroral lines (\SII$\lambda$$\lambda$4068,4076, \OIII$\lambda$4363, \NII$\lambda$5755, \SIII$\lambda$6312, and \OII$\lambda$$\lambda$7320, 7330) when available. 

For our analysis we selected only the sources where the fluxes of all bright and auroral emission lines are measured with $S/N> 5$, resulting in a sub sample of 24 HII regions. %{This choice of $S/N$ threshold for the auroral lines is meant to reduce the uncertainty of the metallicity determination. Indeed it will be shown in Sec. \ref{sec:milyway} how relaxing this constrain can result in  metallicities less constrained to a level which depends on the accuracy of the observed lines. This fact must be taken into account when analysing, e.g., galaxy spectra which might have lower $S/N$ ratios and will be further explored in forthcoming paper dedicated to the measurement of galaxy metallicities (Amiri et al. and other papers in preparation).}
{The choice of S/N threshold for auroral lines aims to reduce uncertainty in metallicity determination. Section \ref{sec:milyway} will demonstrate how relaxing this constraint can lead to less precise metallicities, the degree of which depends on the accuracy of observed lines. This is crucial when analyzing galaxy spectra with potentially lower signal-to-noise ratios. We will explore this further in a forthcoming paper dedicated to measuring galaxy metallicities (Amiri et al., 2024 in preparation).}

We performed model fitting with both dusty and non-dusty clouds - i.e. using a weigthed combination of single-cloud photoionization models either with or without dust mixed to the gas - and with all the ionizing continua presented in section \ref{sec:model}. Each model set $M$ is then characterized by the ionizing continuum and by whether there is dust mixed with the gas, and consists of $U,\NH$ grids of single-cloud models, each computed for a given metallicity $Z$. For each model set $M$, we then find the minimum value of \LF\ by varying $Z$, which we label $\LF_{M,min}$. We then consider $\LF_{min}$ as the smallest of all $\LF_{M,min}$ values, selected by having scaling factors for N and S abundances within $\pm 0.3$ and $\pm 0.15$ dex, respectively. 
Finally, we consider acceptable all the models for which 
\begin{equation}
\LF \le \LF_{min}+0.25
\end{equation}
Following equation \ref{eq:discrepancy}, the increase on average "discrepancy" between model and observations depends on $\varepsilon_a$ and on $\varepsilon_r$ of  the best model with $\LF = \LF_{min}$. 
For instance, given $\varepsilon_a=0.1$, models with $\LF = \LF_{min}+0.25$ have real "discrepancies" $\varepsilon_r=0.05, 0.07, 0.11$
if the best models have $\varepsilon_r=0.01, 0.05, 0.1$, respectively. 
{The choice of the $\LF_{min}+0.25$ threshold is arbitrary and does not affect the best estimate of metallicity determined by the models, as this depends solely on $\LF_{min}$. Increasing the acceptability threshold only widens the range of model metallicities considered consistent with the data.}

\subsection{Observed vs Model Lines\label{sec:obsvsmodel}}
We now compare models and observations by comparing the observed and model line fluxes.
{The left panel of Figure \ref{fig:zurita_chisq}} shows the distribution of the $\LF_{min}$ values for all the HII regions in the selected subsample. We distinguish the cases where, for each HII region, $\LF_{min}$ is found considering all models, only  
models with dust, and those without. When considering models with or without dust, we find $\LF_{min} < 1.1$ except for three cases each where $\LF_{min}$ is $\sim2$ or larger. However, when finding $\LF_{min}$ over all models we always have $\LF_{min} < 1.1$ and the best fit models are given in 20 out of 24 sources by dusty models, while in the rest by models without dust. Indeed, {the left panel of Figure \ref{fig:zurita_chisq}} shows that, overall, dusty models provide in general lower $\LF_{min}$ values.
The distributions of $\LF_{min}$ values thus imply that our multi-cloud models are able to reproduce \textit{all} emission lines with an accuracy better than 10\%; this is an extremely good achievement very rarely matched in previous works (see, e.g., \citealt{chevallard:2018}, \citealt{blanc:2015}, \citealt{morisset_photoionization_2016-2}, \citealt{DAgostino:2019}, \citealt{mingozzi:2020}, \citealt{Olivier:2022}, \citealt{Fernandez:2022}).
It is important to mention that only a small fraction of the single-cloud models in a $8\times 9$ $(U, \NH)$-grid have non zero weights:
in 21 out of 24 best fit models, the numbers of clouds with non-zero weights is between 4 and 9, with only three models being made of 11 or 12 clouds.

The performance of the model with each line is shown in {the right panel of Figure \ref{fig:zurita_chisq}} where we present the distributions of the ratios between observed and model line fluxes for all the selected HII regions and all the models with $\LF\le \LF_{min}+0.25$. The HOMERUN model is able to reproduce to within 10\% both strong and auroral lines, from high (\OIII), intermediate (\SIII) and low excitation (\OII, \SII, \NII) lines. In particular, the model can reproduce both \SII\ and \SIII\ lines at variance with what is found in other studies as already mention in Sec. \ref{sec:obsvsmodel}. In our case \SIII\ lines are those for which we observe the largest discrepancies which are still below 20\% for \SIII$\lambda$9531 in all models while only 11\% of the models present discrepancies larger than 20\% for \SIII$\lambda$6312. 

\begin{figure*}
   \centering
   \includegraphics[width=0.4\linewidth]{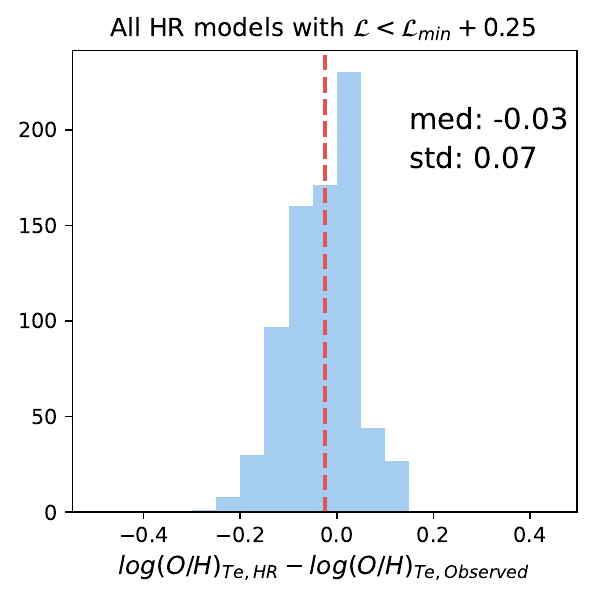}
   \includegraphics[width=0.4\linewidth]{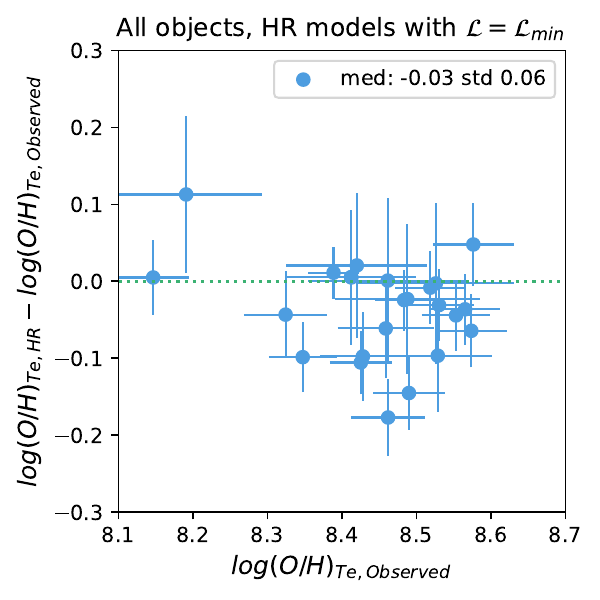}
      \caption{Comparison between the Oxygen abundances derived with the \Te\ method using observed and HOMERUN model line fluxes from the selected HII regions from the sample of \cite{zurita:2021}.
      In the left panel, we consider, for all sources, all the models with $\LF < \LF_{min}+0.25$; the vertical dashed line represents the median of the distribution. In the right panel we consider only the models with $\LF = \LF_{min}$.
      } 
              
         \label{fig:zurita_pyneb}
\end{figure*}

\subsection{Observed vs Model \Te\ metallicities}
We now compare models and observations by comparing the \Te-metallicities obtained by both observed and model line fluxes. We estimated \Te-metallicities using the python package \texttt{pyneb}, which computes the physical conditions and ionic and elemental abundances of photoionized gas by solving the equilibrium equations for an n-level atom \citep{pyneb:2015}.
We selected the same sets of atomic parameters used in CLOUDY and we considered a two-zone cloud where the density of both zones is given by \Ne(S$^{+}$),  the temperature of the low-ionization zone is given by \Te(N$^{+}$) and that of the high-ionization zone by \Te(O$^{+3}$). 
The choice of \Te(N$^{+}$) for the temperature of the low-ionization zone is justified by the fact that \Te(O$^{+}$) can lead to underestimated \Te-metallicities and is the same adopted by \cite{zurita:2021} (see also \citealt{perez-montero:2003, mendez-delgado_density_2023}).
In particular, we adopt the following procedure:
\begin{itemize}
\item \Te(O$^{+3}$) and \Ne(S$^{+}$) are computed simultaneously by combining the \OIII$\lambda$5007/$\lambda$4363 and \SII$\lambda$6731/$\lambda$6716 ratios with the \texttt{getCrossTemDen} procedure of \texttt{pyneb}.
The \SII\ flux ratio is not an accurate tracer of density below the low-density limit (less than a few $\times 10\, \mathrm{cm}^{-3}$) which characterises most HII regions (e.g., \citealt{Nagao:2006}); however the temperature sensitive line ratios that we used are almost independent of density in this regime.
\item we used \Ne(S$^{+}$) to compute \Te(N$^{+}$) from \NII$\lambda$5755/$\lambda$6584, \Te(S$^{+}$) from \SII$\lambda$4069/$\lambda$6720,  \Te(S$^{+2}$) from \SIII$\lambda$6312/$\lambda$9531 and  \Te(O$^{+}$) from \OII$\lambda$7325/$\lambda$3727;
\item we used \Te(N$^{+}$) to compute \Ne(O$^{+}$) from \OII$\lambda$3726/$\lambda$3729;
\item $[\mathrm{O}^{+}/\mathrm{H}]$ and $[\mathrm{O}^{+2}/\mathrm{H}]$ abundance ratios are computed from \OII$\lambda$3727/\Hb, using \Te(N$^{+}$),  \Ne(S$^{+}$), and from \OIII$\lambda$5007/\Hb\ line ratios, using \Te(O$^{+2}$), \Ne(S$^{+}$); $[\mathrm{O}/\mathrm{H}]$ is then the sum of the abundances of the two ions, neglecting any contribution from either neutral Oxygen or $\mathrm{O}^{+3}$ which are typically negligible for HII regions (e.g. \citealt{zurita:2021}).
\item $[\mathrm{N}/\mathrm{O}]$ and $[\mathrm{S}/\mathrm{O}]$ abundance  ratios are similarly computed from \NII$\lambda$6584/\OII$\lambda$3727 and \SII$\lambda$6720/\OII$\lambda$3727 line ratios. Here we assume that in the regions emitting these lines, N, S and O are mostly single ionized ions.
\end{itemize}
Errors are estimated with 100 Montecarlo realisations where the observed line sets are replicated by adding to line fluxes random numbers extracted from a Gaussian distribution with 0 mean and standard deviation equal to the observed relative error (i.e., equal to the assumed discrepancy of $\varepsilon = 10\%$). 
Figure \ref{fig:zurita_pyneb} (left) compares the $T_e$ metallicities derived from the observed line fluxes to the ones predicted by the models: considering all acceptable models ($\LF<\LF_{min}+0.25$) for all objects, the agreement is  good as model and observed metallicities have a median difference of -0.03 dex with a 0.07 dex standard deviation. Similarly, as shown in Figure \ref{fig:zurita_pyneb} (right), when considering only the best models ($\LF=\LF_{min}$) the median difference is -0.03 dex with a 0.06 dex standard deviation. These values are consistent with the $<$10\% discrepancy between model and observed line flux that we required for the HOMERUN model and provide a different measure of the (dis)agreement between models and observations. 
Fig. 3 of \cite{chevallard:2018} shows a similar agreement between O/H from Te and (single cloud) models.
\begin{figure}
   \centering
   \includegraphics[width=0.8\linewidth]{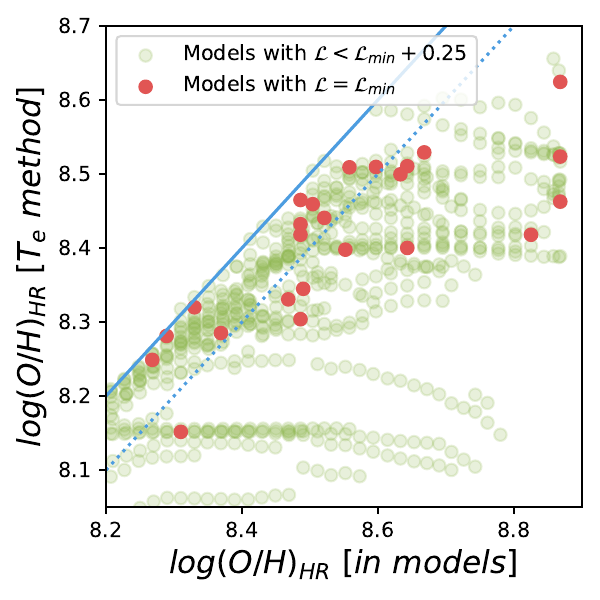}
      \caption{Comparison between the model metallicities inferred from the multi-cloud modelling of the \citealt{zurita:2021} sample (x axis) and the \Te\ metallicities derived from the line fluxes predicted by the same models (y models). The red dots represent the best models for each HII region of the sample, while the green dots represent all "acceptable" models, i.e. those with $\LF< \LF_\mathrm{min}  +0.25$. The solid blue line represents the 1:1 relation, while the dotted blue line represent the case where \Te\ metallicities are 0.1 dex lower than  model metallicities. } 
              
         \label{fig:zurita_model_pyneb}
\end{figure}

Figure \ref{fig:zurita_model_pyneb} presents the comparison between the metallicities used as input of the HOMERUN model calculations, labeled as "$log(O/H)_{HR}~[in~models]$", and the \Te\ metallicities derived with pyneb from the line fluxes predicted by the models, labels as "$log(O/H)_{HR}~[T_e~method]$".  This is not a comparison between models and observations, as it only involves model quantities, but provides an indication on how (non-)accurate the \Te-method is in retrieving the metallicity of a system. The \Te\ metallicities are always smaller than the true values and can be discrepant up to $\sim 0.5$ dex: the median discrepancy and standard deviation of the difference between \Te\ and model metallicities considering only the best models are -0.09 and 0.07 dex, respectively. These values increase to -0.13 and 0.14 dex when considering all models with $\LF<\LF_{min}+0.25$. We believe that this discrepancy is due to the simplified assumptions underlying the \Te\ method (e.g., 2-3 zones with constant density and temperature), confirming the findings by other authors (see, e.g., \citealt{esteban:2014, cameron:2022}), and showing that, with the computational capabilities and the quality and amount of data now available, it is important to adopt more accurate methods for abundance determinations from emission line ratios. A companion paper by Amiri et al. (2024, in preparation) will address in more details the accuracy problems of the \Te\ method.

   \begin{figure*}
   \centering
   \includegraphics[width=0.33\linewidth]{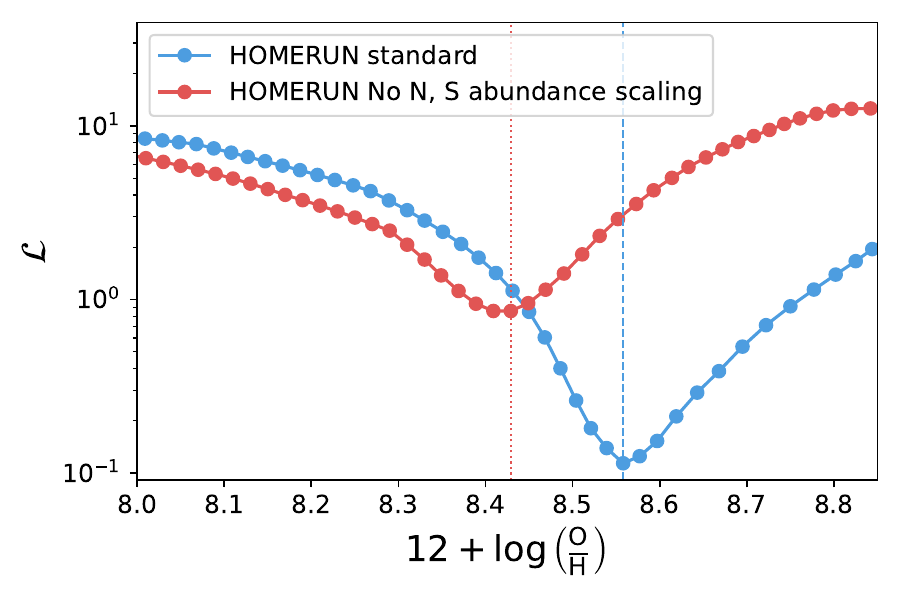}
   \includegraphics[width=0.33\linewidth]{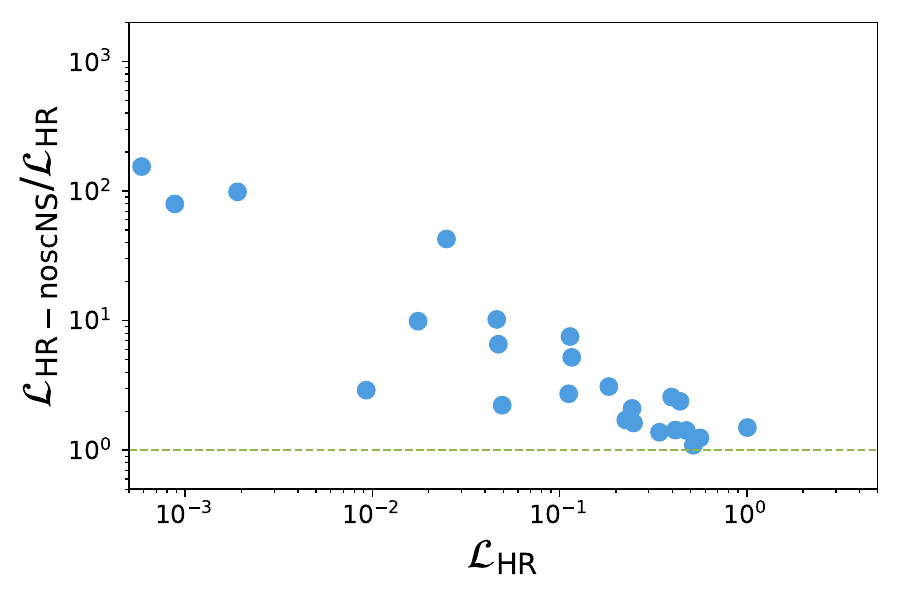}
   \includegraphics[width=0.33\linewidth]{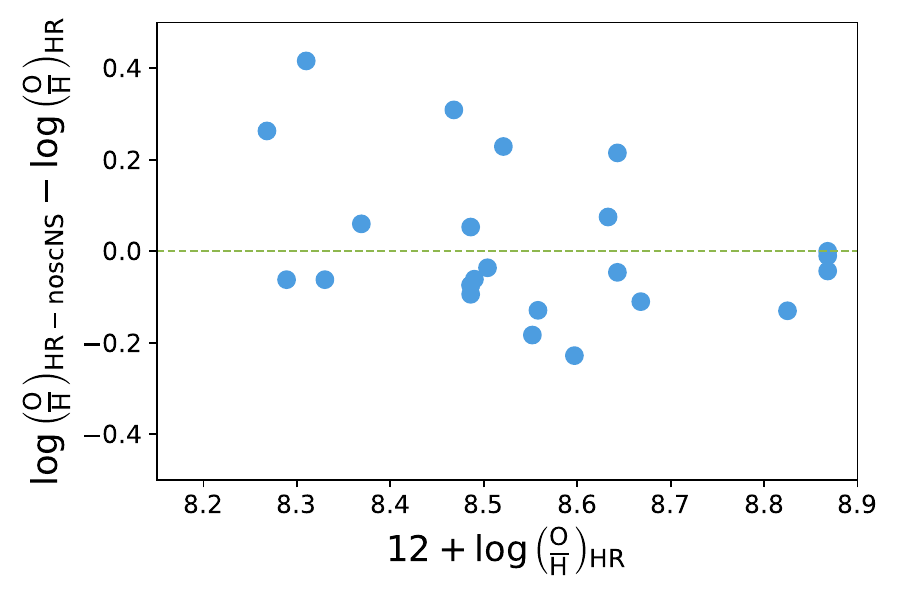}
      \caption{The effect of rescaling N/O and S/O relative abundances with respect  to the values used in photoionization calculations. Left: Variation of the loss function \LF\ as a function of metallicity when fitting the emission line spectrum of the test HII region in Figure 1. The blue dots/line represent the  HOMERUN model with N and S rescaling (standard model), while the red dots/line represent the same model but without rescaling (No N,S abundance scaling). Note that the minimum value of the loss function when excluding re-scaling is $\sim$ten times higher than for the standard model. The red vertical dashed and dotted lines represent the metallicity of the models with the minimum value of the loss function \LF.
 Middle: ratios of \LF\ minimum values from models with no N,S abundance scaling and standard models when applied to the HII regions of the Zurita et al. (2021)
sample. The effect of non rescaling N and S is to increase the minimum \LF\ value by up to a factor 100.
  Right: variation of abundances of models with no N,S abundance scaling with respect to standard models compared to the abundances of standard models.
      }
              
         \label{fig:model_noscalingNS}
   \end{figure*}

\subsection{The effect of varying N/O and S/O relative abundances}

Figure \ref{fig:model_noscalingNS} outlines the importance of allowing for a scaling factor of the N and S abundances. The left panel shows how in the example model the minimum value of the \LF\ function increases by a factor $\simeq 10$ when no abundance scaling is taken into account. Overall, when fitting all the HII regions in the \cite{zurita:2021} sample (see Section \ref{sec:zurita}),  minimum \LF\ values increase by factors up to over 100 (center/middle panel), while, at the same time, metallicities are over- or under-estimated by up to 0.4 dex in case of no abundance scaling (right panel). Clearly,  rescaling the N and S abundances with respect to the ones assumed in the photoionization computations is extremely important to improve the comparison of models and observations and the reliability of the metallicity estimates (e.g., \citealt{strom_measuring_2018}).

  \begin{figure*}
   \centering
   \includegraphics[width=0.33\linewidth]{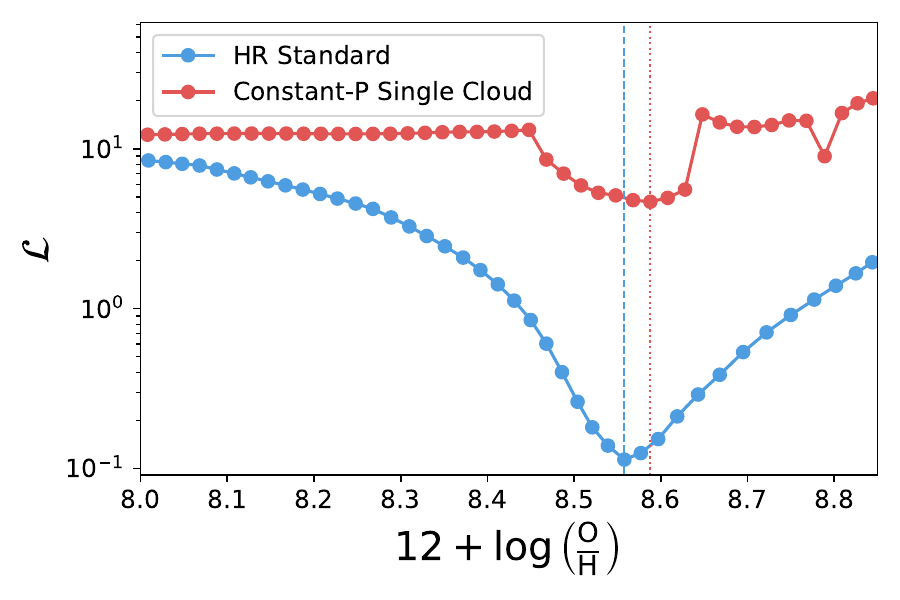}
   \includegraphics[width=0.33\linewidth]{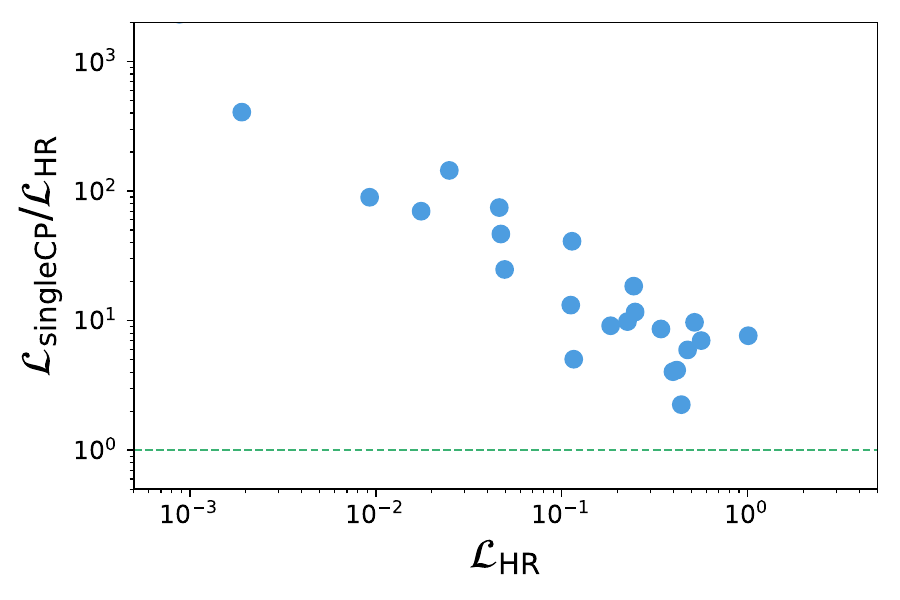}
   \includegraphics[width=0.33\linewidth]{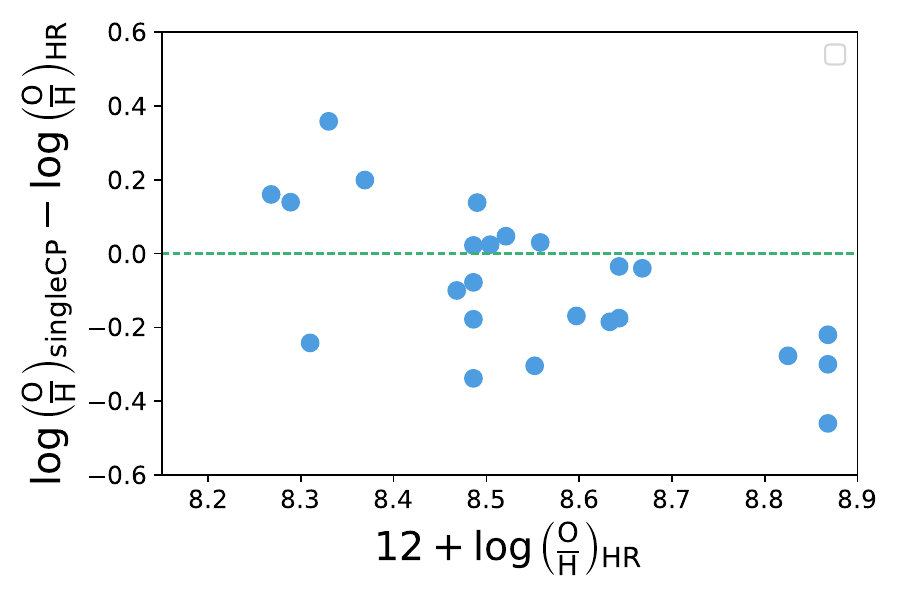}
      \caption{Comparison of HOMERUN with single cloud, constant pressure models. Left: Variation of the loss function \LF\ as a function of metallicity for the emission line spectrum of the test HII region. The blue dots/line represent the HOMERUN model presented here (standard model), while the orange dots/line represent the case of constant pressure single cloud models (Constant-P single cloud). The red vertical dashed lines represent the metallicity of the models with the minimum value of the loss function \LF.
 Middle: ratios of \LF\ minimum values from Constant-P single cloud  and HOMERUN standard models. 
  Right: variation of abundances of Constant-P single cloud  models with respect to HOMERUN standard models. 
    }
              
         \label{fig:model_singlecloud}
   \end{figure*}

\subsection{Comparison with constant-pressure single-cloud models \label{sec:literature}}

Figure \ref{fig:model_singlecloud} shows the large improvement of our multi-cloud approach with respect to the case of the single cloud constant pressure models which are used in the literature by many authors (e.g., \citealt{blanc:2015,dopita:2016,perez-montero:2021} and references therein). 

The left panel compares the \LF\ curves obtained with the single ({red}) and multi-cloud (blue) approach applied to the  test HII region from \cite{zurita:2021} used before. The blue curve is the same as in Figures \ref{fig:model_example} and \ref{fig:model_noscalingNS}.

In order to obtain the orange curve, we have obtained grids of constant-pressure, single-cloud models following the same approach as before: for each given ionising continuum and gas metallicity, we have computed a grid of models with varying $U$ and $\NH$. For each single-cloud model in the grid we have explored the same ($U$, \NH) grid used in the multi-cloud model to determine the optimal ($U$, \NH) pair that minimises the loss function, along with the best scalings for N and S abundances, at given gas metallicity and ionising continuum.  We have then considered the single cloud model with the lowest \LF\  value.
In practice, we have  selected the single-cloud, constant-pressure model which provides the minimum value of the loss function for a given continuum, gas abundance, ionization parameter and density. The difference with respect to other models in the literature is that we allow for a variation of N/O and S/O and/or we consider a much wider grid in density (e.g., \citealt{van-Zee:1998, dopita_new_2013,  blanc:2015,strom_measuring_2018, Yan:2018}).

 In the case of this particular example, the minimum \LF\ value of our multi-cloud approach is more than a factor ten lower although the estimated gas metallicity is similar. The comparison between the orange curves in Figures \ref{fig:model_example} and \ref{fig:model_noscalingNS} shows that rescaling N and S abundances is crucial for having a low \LF\ minimum value. Indeed, the single-cloud constant pressure model has a similar $\LF_{min}$ value as the multi-cloud model where the N and S abundances are those assumed in  the photoionization calculations.

The center and right panels in the figure compare the minimum \LF\ values and the oxygen abundances obtained by applying the two methods to the HII regions of the \cite{zurita:2021} sample: \LF\ values obtained with  the multi-cloud approach are more than a factor 10 lower, but the discrepancy can increase by two orders of magnitude.  On the other hand, discrepancies in oxygen abundances are less severe and are up to $\pm 0.3-0.4$ dex.  

Finally, we note that the difference between the ionization parameter of the single-cloud, constant-pressure models and the average $U$ of the HOMERUN models is on average -0.2 dex with a standard deviation of 0.5 dex. Similarly, the differences in densities are on average 0.2 dex but with a standard deviation of 1.2 dex.

\section{The sample of HII regions from the CHAOS project\label{sec:berg}}
\begin{figure*}
  \begin{minipage}[!t]{0.35\textwidth}
    %\centering
    \includegraphics[width=\linewidth]{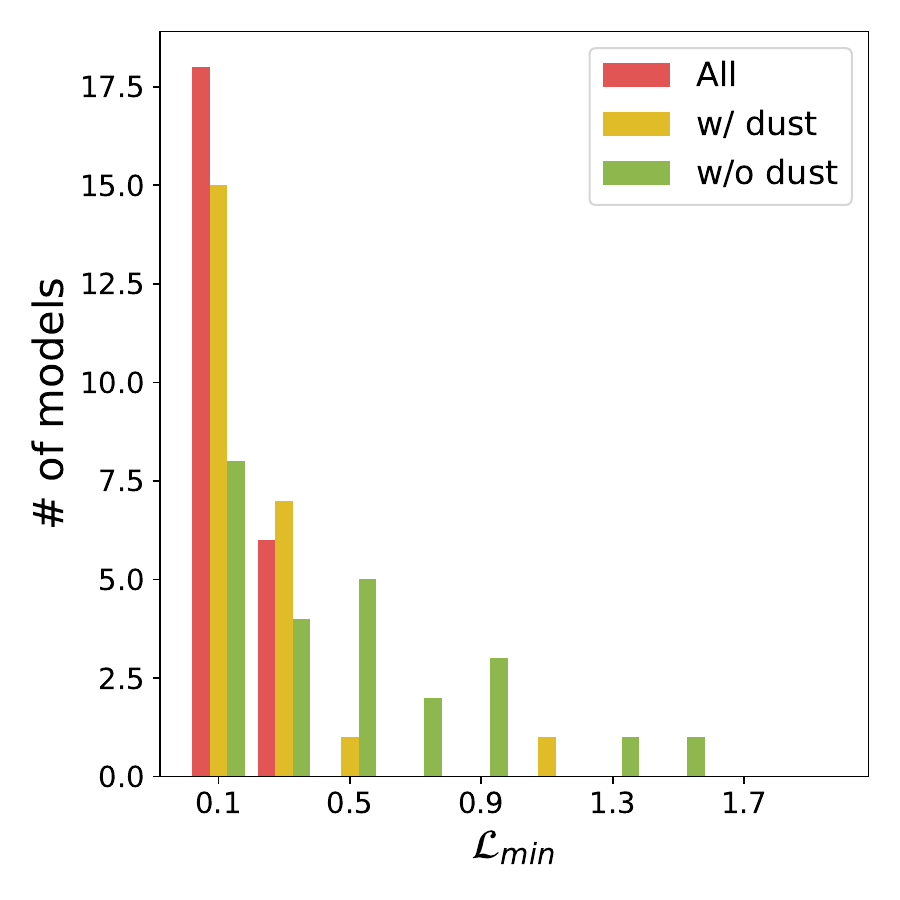}
  \end{minipage}
  \begin{minipage}[!t]{0.64\textwidth}
    %\centering
    \includegraphics[width=\linewidth]{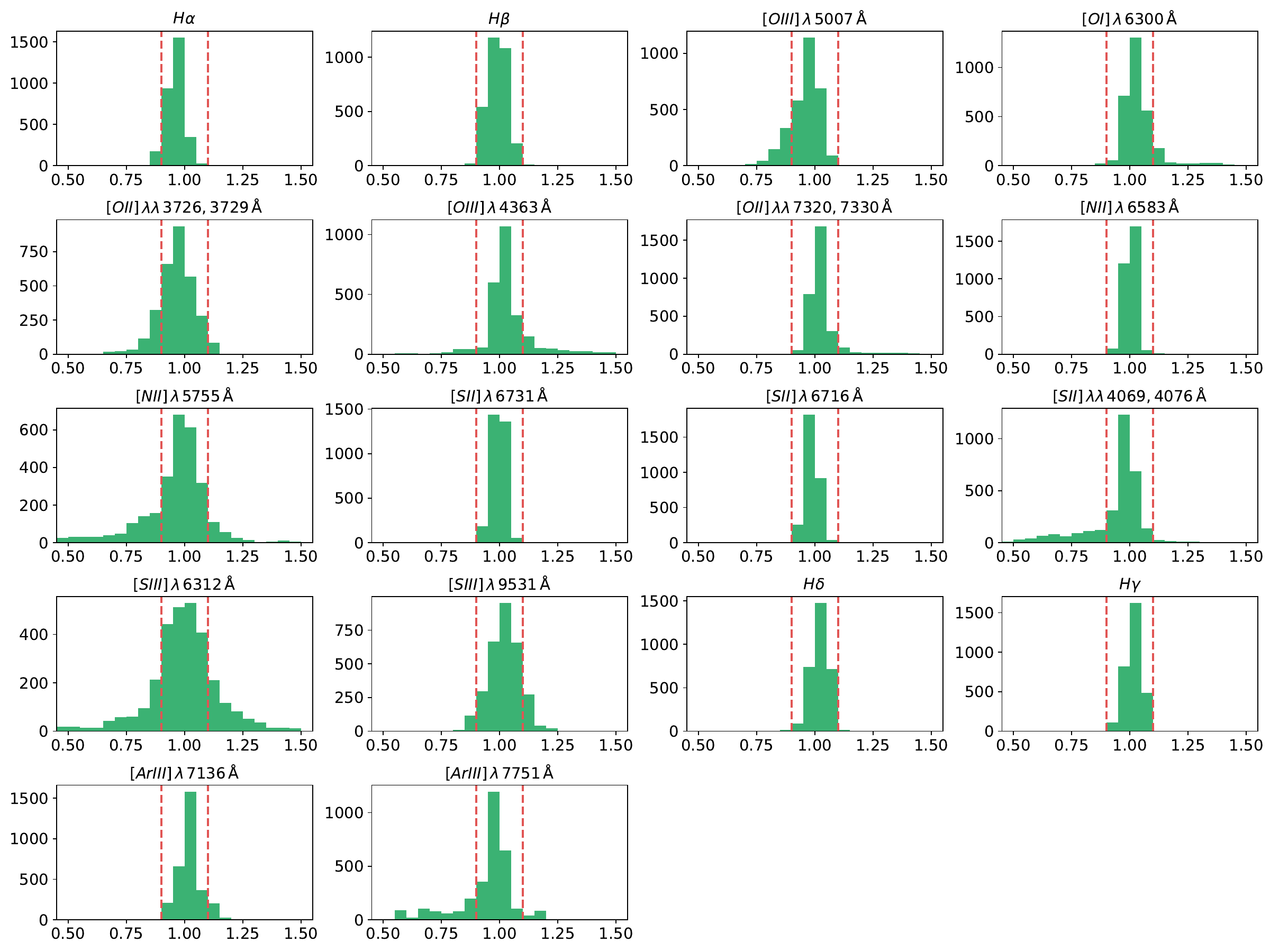}
  \end{minipage}

     \caption{Left: distribution of \LF\ values for the best/acceptable models of the HII regions from CHAOS sample. Right: distributions of the contributions to \LF\ from each of the lines used for model fitting (the adopted error on line fluxes is the maximum between the observed error and 10\%). } 
              
         \label{fig:berg_chisq}
\end{figure*}
\begin{figure*}
   \centering
   \includegraphics[width=0.45\linewidth]{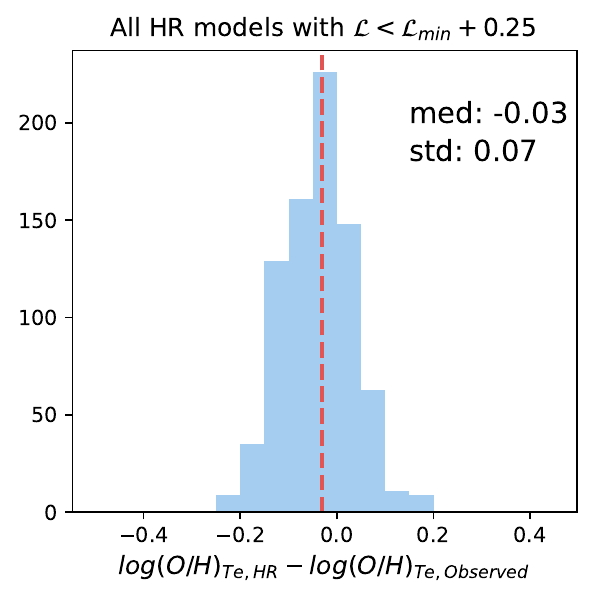}
   \includegraphics[width=0.45\linewidth]{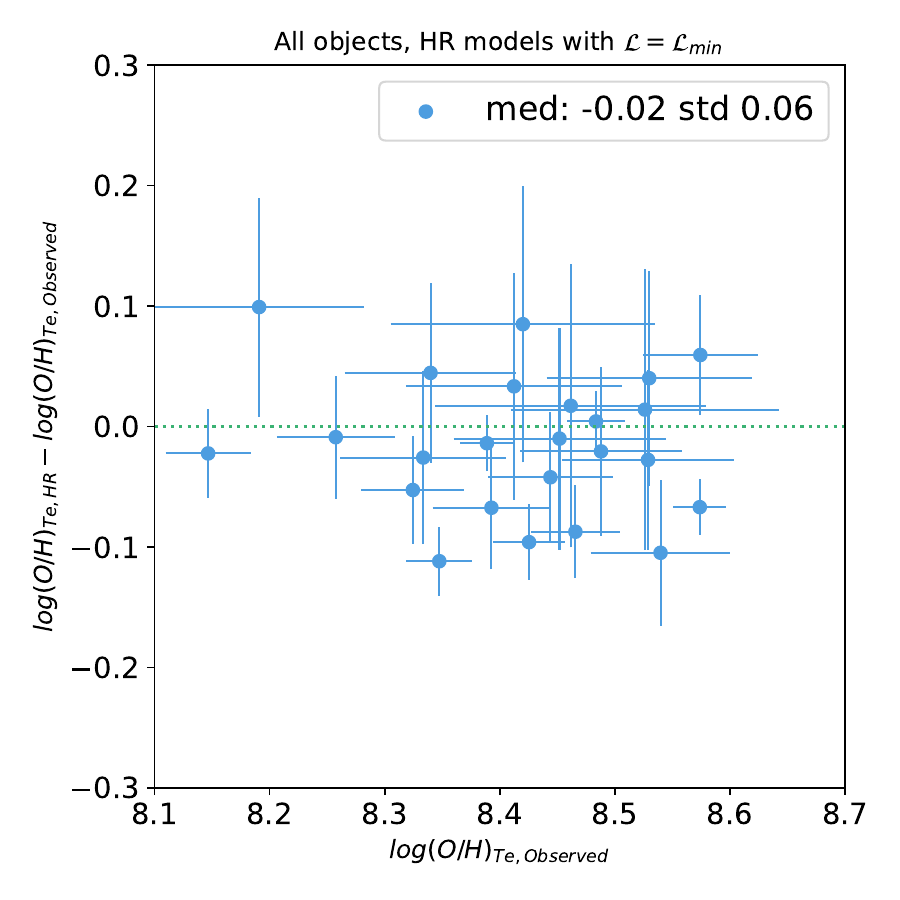}
      \caption{Comparison between the Oxygen abundances derived with the \Te\ method using observed and HOMERUN model line fluxes from the selected HII regions from the CHAOS sample. In the left panel, we consider, for all sources, all the models with $\LF< \LF_{min}+0.25$; the vertical dashed line represents the median of the distribution. In the right panel we consider only the models with $\LF = \LF_{min}$.  } 
              
         \label{fig:berg_pyneb}
\end{figure*}
\begin{figure}
   \centering
   \includegraphics[width=\linewidth]{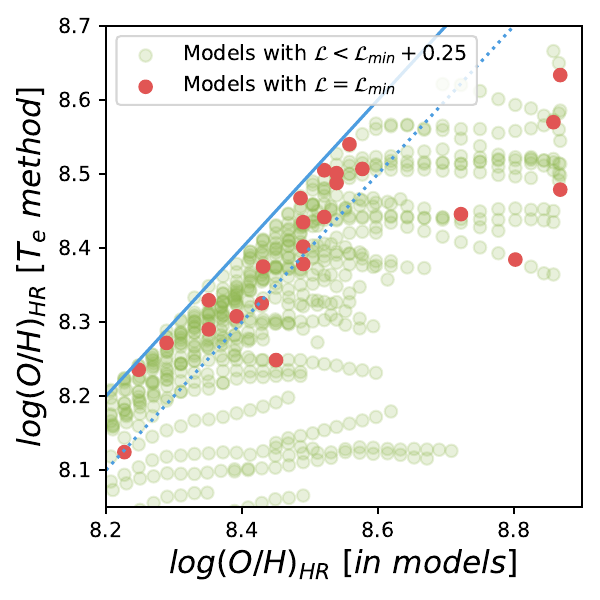}
      \caption{Comparison between the model metallicities inferred from the multi-cloud modelling of the CHAOS sample (x axis) and the \Te\ metallicities derived from the line fluxes predicted by the same models (y models). The red dots represent the best models for each HII region of the sample, while the green dots represent all "acceptable" models, i.e. those with $\LF< \LF_\mathrm{min}  +0.25$. The solid blue line represents the 1:1 relation, while the dotted blue line represent the case where \Te\ metallicities are 0.1 dex lower than  model metallicities.
} 
              
         \label{fig:berg_model_pyneb}
\end{figure}

We now apply the model to a sample of HII regions with more homogeneous data, higher S/N and a larger number of emission lines.

We consider the HII regions of six galaxies from the CHemical Abundances of Spirals (CHAOS) project (\citealt{chaosI,chaosII,chaosIII,chaosIV,rogers:2021,rogers:2022}) and we apply the same procedure followed in the previous section. At variance with the compilation by \citealt{zurita:2021}, the data now include 11 additional lines (see figure \ref{fig:berg_chisq} for the list) including  \OI$\lambda$6300, a line which has always been difficult to account for in photoionization models (e.g., \citealt{dopita_new_2013}). We select the HII regions with the same criterion used for the \cite{zurita:2021} sample, i.e. the auroral lines detected with $S/N>5$.

Figures \ref{fig:berg_chisq} and \ref{fig:berg_pyneb} confirm the results of the previous section: the $\LF_\mathrm{min}$ values of the best models are all lower than 1 (red bars in figure) indicating an average agreement between model and observed line fluxes better than 10\%. All lines are well reproduced and in particular, the models are able to reproduce at the same time \OI, \OII\ and \OIII, as well as \SII\ and \SIII, which is a significant improvement compared to previous works (see, e.g., \citealt{mingozzi:2020} on \SIII\ and \citealt{dopita_new_2013} on \OI)
Note that the [NeIII]$\lambda$3869 line does not provide any constraint to the model since it is the only Neon line, and it is used only to estimate Ne abundance.
It is remarkable that when using only the emission lines from the \cite{zurita:2021} sample and \OI$\lambda$6300, the $\LF_\mathrm{min}$ values are all lower than 0.3. 
Similarly to the case of the \cite{zurita:2021} sample, only a small fraction of the single-cloud models in
a 8 × 9 (U, NH)-grid have non zero weights: in 23 out of 24 best
fit models, the numbers of clouds with non-zero weights is {between} 5 and 10, with only one models being made of 12
clouds.

We have estimated $T_e$ metallicities as in the previous section and found that the agreement between estimates from observed and model line fluxes is similar to what found for the \cite{zurita:2021} sample (figure \ref{fig:berg_pyneb}). Finally,
 we confirm as well the discrepancy of up to 0.2 dex between the metallicities used in the photoionization model computations and the $T_e$ metallicities derived from the line fluxes predicted by the same models (figure \ref{fig:berg_model_pyneb}). 

Outside of \HII\ regions, the warm ($\Te\sim 10^4\,\mathrm{K}$), ionized component of the ISM is made of a more diffuse medium, the so-called DIG, particularly evident above and below the galactic plane. Such medium can significantly affect the integrated spectra of \HII\ regions and star forming galaxies, depending on the size of the aperture from which spectra are extracted.
\cite{mannucci:2021} investigated the 
 contribution of the DIG which is often invoked to explain the differences between the spectra of HII regions in local galaxies and the total or kpc-scale spectra of star-forming galaxies when the flux ratios involving low-ionization emission lines are considered (e.g., \citealt{belfiore:2016, Kumari:2019, Vale-Asari:2019, Della-Bruna:2020, Sanders:2020}). Comparing  single slit spectra of HII regions from CHAOS with those extracted with different aperture sizes from integral field spectra, they concluded that these differences are mostly ascribed to the small anglular size of the slits, which is not covering the full extent of the HII region, rather than to the DIG contamination. In principle, our model approach, based on combining multiple clouds to reproduce the correct volume of gas occupied by any given ion \Ioni\ with the correct \Te\ and \Ne, can naturally take into account both aperture effects and the contributions from regions with different physical properties like the DIG. However, this issue will be addressed in a forthcoming paper. 

\section{Gas and stellar metallicities in the Milky Way\label{sec:milyway}}

\begin{figure*}
\centering
   \includegraphics[width=0.9\linewidth]{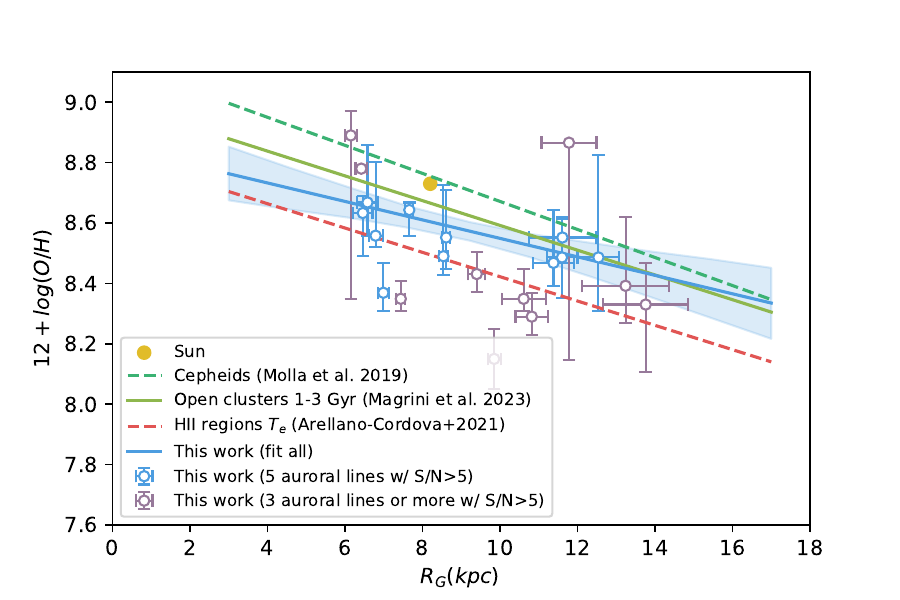}
       \caption{Comparison between our metallicity estimates of HII regions in our Galaxy, the metallicity gradient from the \Te\ abundances of HII regions (\citealt{arellano-cordova:2021}), the Cepheids (\citealt{molla:2019}) and intermediate age open clusters (\citealt{magrini_gaia-eso_2023}). The blue line surrounded by the blue shaded area represents the linear fit, with uncertainties, to all our metallicity estimates.} 
              
         \label{fig:milkyway_gradient}
\end{figure*}

We present the last application of our method and its potential to solve the observed discrepancy between the metallicities of HII regions and those of the stars in our Galaxy.

The metallicity of the ionized gas is generally {in} good agreement with that of the stars, at least in external galaxies (see, e.g., Fig. 11 of \citealt{Bresolin:2016}). However, 
there is a discrepancy between the metallicities of HII regions in our Galaxy, derived from the application of the \Te-method to gas emission lines, and the stellar metallicities derived from B-type stars, Classical Cepheids and Open clusters, which are $\sim 0.2$ dex larger (e.g., Fig. 4 of \citealt{esteban:2022}). 
Such difference is not present when considering gas-phase metallicities from recombination lines and is part of the so-called abundance discrepancy problem which, as described in Sec. \ref{sec:introduction}, is quite a common occurrence when comparing \Te\ and recombination-line-based metallicities. As mentioned by \cite{esteban:2022}, such discrepancies might be related to temperature fluctuations which bias auroral lines towards the hotter parts of the HII regions providing higher than average temperatures and consequently lower abundances \citep{peimbert:1967}. Indeed, this discrepancy disappears when correction for temperature fluctuations and ionization fractions are applied \citep{mendez-delgado:2022}. Still corrections for temperature fluctuations are difficult to estimate and ionization corrections must be derived from photoionization models, thus adding up additional assumptions and uncertainties to the \Te\ method.

We consider all the Galactic HII regions from table 1 of  \citep{mendez-delgado:2022} which have at least three of the five auroral lines considered so far detected with $S/N>5$ and estimate their metallicities by applying our method to the emission lines fluxes from \cite{arellano-cordova:2020}, \citeyear{arellano-cordova:2021} and \cite{zurita:2021}. 
Figure \ref{fig:milkyway_gradient} shows the HII region gas-phase metallicities plotted as a function of GAIA-based galactocentric distances from  \cite{mendez-delgado:2022}, and compare them with the metallicity gradient of the same HII regions estimated with the \Te-method without any ionization correction \citep{arellano-cordova:2021}, the Cepheid metallicity gradient from \cite{molla:2019} and the intermediate-age (1-3 Gyr) open clusters metallicity gradient from \cite{magrini_gaia-eso_2023}. The linear fit considers all HII regions and was performed with the \texttt{ltsfit}\footnote{Available at \href{https://www-astro.physics.ox.ac.uk/~cappellari/software/}{https://www-astro.physics.ox.ac.uk/$\sim$cappellari/software/}} python package which implements a Robust linear regression with scatter in one or two dimensions \citep{Cappellari:2013}.
{Unlike Sections \ref{sec:zurita} and \ref{sec:berg}, where we required HII regions to have 5 auroral lines with $S/N>5$, we have reduced the selection threshold to 3 auroral lines with $S/N>5$ to expand our sample size. This change, as the figure demonstrates, does not significantly alter our conclusions but may slightly increase the scatter.}
The plot clearly shows how our model-based metallicities are, on average, in agreement with the stellar metallicities especially when considering that oxygen is a refractory element and is likely depleted into stars: indeed, it is important to recall that these are gas-phase metallicities which do not take into account the depletion of metals into dust.
Oxygen depletion can vary between 10-20\% \citep{Psaradaki:2023, amayo_ionization_2021} and the 40\% assumed by CLOUDY. Therefore,  any abundance discrepancy here can be easily explained with a $\lesssim 0.1$ dex depletion of oxygen into dust without having to rely on uncertain temperature fluctuations or ionization corrections.

\section{Summary and Conclusions\label{sec:summary}}

We have presented a new approach to modelling emission lines from photoionized gas which allows to reproduce observed line ratios from a wide range of ionization and with an accuracy to better than 10\%, and whose main application is the accurate determination of gas metallicity.
\begin{itemize}
\item Our approach is based on the weighted combination of multiple single-cloud photoionization models computed with CLOUDY, as in previous works. The novelty of our approach is that the weights of the single-cloud models  are not parametric nor assumed a priori, but are free parameters of the fit. 
The resulting best fit model accounts for the varying physical conditions (temperatures, densities, ionisation conditions and gas metallicities) observed along the line of sight of the interstellar gas of HII regions and galaxies.
\item Since  observed line luminosities are integrated quantities over the entire volume of the emitting source, the single-cloud models contributing to a best fit do not necessarily correspond to physical entities but are the basic \textit{building blocks} of the model. Therefore, if a best fit model has a single cloud with specific $U$, \NH, this does not necessarily mean that such cloud exists but that the total line emission has a contribution from gas with those physical parameters, regardless of where and how it is spatially distributed.
\item Our model has as many free parameters as the single-cloud models considered, however the fitting of the observed data is not degenerate and the loss function (\LF) curves, computed as a function of  gas metallicity, have very deep and well defined minima. Indeed our approach is similar to the one used for fitting the stellar continua of galaxies where many hundreds of templates can be combined with weights which are free parameters of the fit (see, e.g., the \texttt{ppxf} code to analyze stellar continuum spectra).
\item We have shown that a critical point for our modelling is to allow for different abundances for elements like N and S than those used in the photoionisation calculations: N and S emission lines are not the major coolants of the photoionised region like O lines and, therefore, their luminosities scale almost linearly with gas abundances.  Not accounting for N and S abundance variations can significantly affect the quality of the fit (with minimum \LF\ values worse by even a couple of orders of magnitude) and the estimated gas metallicity. 
\item We have also shown that our approach provides a significant improvement compared to the single-cloud, constant-pressure models commonly used in the literature, as well as being able to model a larger set of emission lines. Indeed, several works based on single-cloud photoionization modelling claimed to pin down the metallicity with high accuracy by using just a small set of lines, typically the optical strong lines and they rarely, if ever, 
include auroral lines which are instead successfully reproduced in HOMERUN.
\item We have applied our model to the HII regions from the samples by \cite{zurita:2021} and from the CHAOS project. For the sample by \cite{zurita:2021}, we have considered HII regions where all the following lines were detected with $S/N>5$:
\OII$\lambda\lambda$3726,3729, \OIII$\lambda\lambda$4959,5007, \NII$\lambda\lambda$6548,6583, \Ha, \SII$\lambda\lambda$6717,6731, \SIII$\lambda\lambda$9069,9532, \SII$\lambda\lambda$4068,4076, \OIII$\lambda$4363, \NII$\lambda$5755, \SIII$\lambda$6312, and \OII$\lambda\lambda$7320, 7330. All the observed emission lines for the 24 selected HII regions were reproduced by the model to better than 10\%. In particular the model was able to reproduce also the \SIII 9069,9532 lines which are well known to be problematic. The \Te-metallicities derived from the observed emission lines and those derived from the model emission lines agree to within 0.05 dex, another indication of the extremely good performance of the models. 
\item However, we showed that \Te-metallicities suffer from uncertainties which are related to the simplified assumptions underlying the method: the \Te-metallicities computed from the emission lines predicted by the models 
can be more than 0.1 dex smaller than the real gas metallicities of the models themselves. 
\item The CHAOS sample is more homogeneous in terms of data quality, is characterized by better S/N and provides  the fluxes of many more emission lines. Overall, we have modeled the same emission lines used for the  \cite{zurita:2021} sample but we have added also other lines, doubling their total number. 
\item Even with these increased constraints, the models reproduce the observed emission lines to better than 10\%. It is worth noting that the HOMERUN models are able to simultaneously reproduce the emission lines from very different ionization stages, like \OI, \OII\ and \OIII. 
\item Finally, we have shown that the gas metallicities estimated with our models in HII regions of the Milky Way are in better agreement with the stellar metallicities than the estimates based on the \Te-method. 
\end{itemize}
Overall, our method provides a new accurate tool to estimate the metallicity and the physical conditions of the ionized gas. It can be applied in many different science cases and wherever there are emission lines from photoionized gas.

There are many applications of our method which will be presented in a series of forthcoming papers. In particular, the second paper in this series (Amiri et al. 2024, in preparation) will present the application of  this method to the spectra of galaxies, will determine their metallicities and will provide a new self-consistent calibration of the strong line method. 
\textbf{Future studies will explore the limitations and informational content revealed by incorporating ultra-violet, mid-infrared, and far-infrared spectral lines when modeling spectra of HII regions, star-forming galaxies, and Active Galactic Nuclei (AGN). }

\begin{acknowledgements}
We are grateful to Gary Ferland and his collaborators for the development of CLOUDY and for allowing its free use for the community.
This work has been partially financed by the European Union with the Next Generation EU plan, through PRIN-MUR project "PROMETEUS" (202223XPZM).
We acknowledge the support of the INAF Large Grant 2022 “The metal circle: a new sharp view of the baryon cycle up to Cosmic Dawn with the latest generation IFU facilities” and INAF Large Grant 2022 "Dual and binary SMBH in the multi-messenger era".
SC and GV acknowledge support by European Union’s HE ERC Starting Grant No. 101040227 - WINGS.
\end{acknowledgements}

\bibliographystyle{aa} % style aa.bst
\bibliography{bibliography} % your references Yourfile.bib

\end{document}